\pdfoutput=1
\documentclass[preprint,prb,superscriptaddress,floatfix]{revtex4-1}
\usepackage{amsmath,amssymb}
\usepackage[T1]{fontenc}
\usepackage[utf8]{inputenc}
\usepackage{color}
\usepackage{libertine}
\usepackage[libertine]{newtxmath} 
\usepackage[protrusion=true,expansion=true,stretch=10]{microtype}
\usepackage{ulem}
\usepackage[version=4]{mhchem} 
\usepackage{siunitx,physics} 
\usepackage{graphicx}
\usepackage{soul}
\usepackage[nointegrals]{wasysym}
\newcommand*{\twosymb}[2]{\textit{#1}\thinspace--\thinspace\textit{#2}}

\DeclareSIUnit\oe{Oe}
\DeclareSIUnit\pcc{\centi\meter^{-3}}

\begin{document}
\begin{abstract}
	\setlength{\emergencystretch}{2pt}
	The change of a material's electrical resistance~(\textit{R}) in response to an external magnetic field~(\textit{B}) provides subtle information for the characterization of its electronic properties and has found applications in sensor and storage related technologies. In good metals, Boltzmann's theory predicts a quadratic growth in magnetoresistance~(\textit{MR}) at low \textit{B}, and saturation at high fields. On the other hand, a number of non-magnetic materials with weak electronic correlation and low carrier concentration for metallicity, such as inhomogeneous conductors, semimetals, narrow gap semiconductors and topological insulators, two dimensional electron gas (2DEG) show positive, non-saturating linear magnetoresistance~(\textit{LMR}). 
	However, observation of \textit{LMR} in single crystals of a good metal is rare. 
	Here we present low-temperature, angle-dependent magnetotransport in single crystals of the antiferromagnetic metal, \ce{TmB4}. 
	We observe large, positive and anisotropic \textit{MR}(\textit{B}), which can be tuned from quadratic to linear by changing the direction of the applied field. In view of the fact that isotropic, single crystalline metals with large Fermi surface~(\textit{FS}) are not expected to exhibit \textit{LMR}, we attribute our observations to the anisotropic \textit{FS} topology of \ce{TmB4}. Furthermore, the linear \textit{MR} is found to be temperature-independent, suggestive of quantum mechanical origin.
\end{abstract}	
\title{Quadratic to linear magnetoresistance tuning in \ce{TmB4}}
\author{Sreemanta Mitra}
\altaffiliation[Current address: ]{Department of Physics, Indian Institute of Science, Bangalore 560012, India.~}\email{sreemanta85@gmail.com}
\affiliation{Division of Physics and Applied Physics, School of Physical and Mathematical Sciences, Nanyang Technological University, 21, Nanyang Link 637371, Singapore.}
\author{Jeremy Goh Swee Kang}
\affiliation{Division of Physics and Applied Physics, School of Physical and Mathematical Sciences, Nanyang Technological University, 21, Nanyang Link 637371, Singapore.}
\author{John Shin}
\affiliation{Department of Physics, University of California, Santa Cruz, California 95064, USA.}
\author{Jin Quan Ng}
\affiliation{Division of Physics and Applied Physics, School of Physical and Mathematical Sciences, Nanyang Technological University, 21, Nanyang Link 637371, Singapore.}
\author{Sai Swaroop Sunku}
\altaffiliation[Current address: ]{Department of Physics, Columbia University, New York, 10027, USA.}
\affiliation{Division of Physics and Applied Physics, School of Physical and Mathematical Sciences, Nanyang Technological University, 21, Nanyang Link 637371, Singapore.}
\author{Tai Kong}
\altaffiliation[Current address: ]{Department of Chemistry, Princeton University, New Jersey, 08544, USA.}
\affiliation{Ames Laboratory, U.S. DOE and Department of Physics and Astronomy, Iowa State University, Ames, Iowa, 50011, USA.}
\author{Paul C. Canfield}
\affiliation{Ames Laboratory, U.S. DOE and Department of Physics and Astronomy, Iowa State University, Ames, Iowa, 50011, USA.}
\author{B.~ Sriram Shastry}
\affiliation{Department of Physics, University of California, Santa Cruz, California 95064, USA.}
\author{Pinaki Sengupta}
\author{Christos Panagopoulos}
\email{christos@ntu.edu.sg}
\affiliation{Division of Physics and Applied Physics, School of Physical and Mathematical Sciences, Nanyang Technological University, 21, Nanyang Link 637371, Singapore.}
\maketitle

\section{Introduction}
Interest in novel magnetotransport phenomena in metallic magnets is driven by technological and fundamental considerations. The technological motivation comes
from harnessing the unique functionalities associated with properties such as giant magnetoresistance, while the fundamental motivation arises from discovering and understanding new quantum many body physics. The quest for linear magnetoresistance (\textit{LMR}) in strongly correlated systems is one such example of fundamental motivation\cite{cras}.
Boltzmann's classical electronic transport theory shows that in a conductor with a large Fermi surface
(\textit{FS}), magnetoresistance, \textit{MR} (defined as $\frac{\Delta\rho(B)}{\rho(0)}=\frac{\rho(B)-\rho(0)}{\rho(0)}$, where $\rho(B)$ is resistivity in magnetic field {\textit{B}}) grows as $B^2$ at small fields and saturates to a constant value at higher fields\cite{pip}. A linear and non-saturating dependence on $B$ denotes a departure from conventional behavior. Notably, \textit{LMR} has been found to arise from multiple factors ranging from classical\cite{parnat,parprb,apl,prl2015,dsm2,justin,prb95,prl117} to quantum\cite{ab1,ab2}. Discovery and understanding of \textit{LMR} in new materials, and controlling the underlying mechanism remains an active research frontier\cite{dsm,biSc,litnat, parnat,parprb,apl,graphi,prl2015,justin,prb95,prl117,dsm2,graNL,caprb, sinat,natmat,TI0,TI1,natphy16,cras,ab1,ab2,ab3,zhaoprb,amirag}.  
   
The super-linear, non-saturating \textit{MR} observed in non-stoichiometric silver chalcogenides\cite{litnat} (\ce{Ag_{2+$\delta$}Se, Ag_{2+$\delta$}Te}), 2DEG\cite{amirag}, \ce{Bi2Se3}\cite{apl} were explained using a classical random-resistor model\cite{parnat,parprb}. Mobility ($\mu$)\cite{dsm2} and density\cite{prl117} fluctuations, along with space-charge effect\cite{sinat}  have also been discussed to be the primary origin of \textit{LMR} in several materials. 
On the other hand, \textit{LMR} in single crystals of semimetals\cite{graNL,caprb,zhaoprb}, narrow gap semiconductors\cite{natmat}, topological insulators\cite{TI0,TI1} and pressure-induced superconductors\cite{cras} have been explained with a quantum picture\cite{ab1,ab2}. 
In single crystalline metals with parabolic dispersion, \textit{LMR} is atypical and only observed previously in some members of the light rare-earth diantimonide (\ce{RSb2}) and \ce{RAgSb2}~[R=La-Nd, Sm] families\cite{redsb,reagsb2}. 
Hence, it would be interesting to explore a metal where not only expected quadratic \textit{MR} is realized, but also a tuning to \textit{LMR} can be achieved  by changing certain experimental parameters, while maintaining the purity and stoichiometry of the single crystal.

\begin{figure}
        \includegraphics[width=0.8\linewidth]{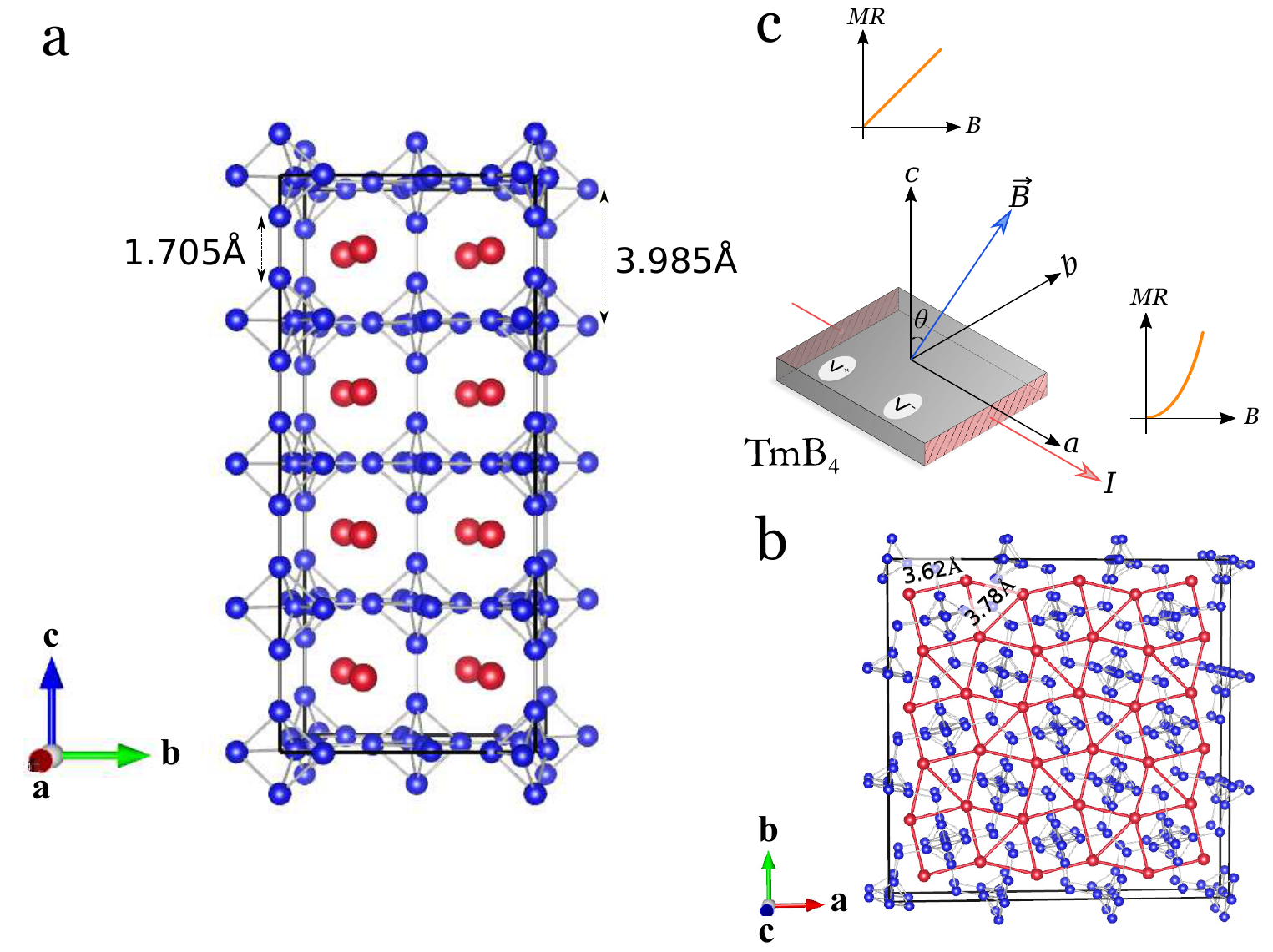}
    	\caption{Structural and experimental considerations. (a)~The alternating-layer crystal structure of \ce{TmB4}. Four unit cells stacked along the \textit{c}-axis are shown. The \ce{Tm} (red) atom planes lie halfway between the \ce{B} (blue) atom layers, which are separated by a distance of \SI{3.985}{\angstrom}. One type of boron lies exclusively within the boron plane, whereas the other type is part of the boron plane and an octahedral chain along the \textit{c}-axis. (b)~A unit cell of \ce{TmB4} viewed along the \textit{c}-axis. The sub-lattice of \ce{Tm} atoms maps to a topologically equivalent Shastry-Sutherland lattice structure\cite{ssl} with perfect squares and nearly equilateral triangles of sides \SI{3.62}{\angstrom} and \SI{3.78}{\angstrom} respectively. The crystal structure of \ce{TmB4} is prepared using VESTA\cite{vesta}. (c)~A schematic of the experimental arrangement and main results. $\theta$ is the tilt angle between \textit{B} and the crystal \textit{c}-axis. The excitation current \textit{I} is applied parallel to the \textit{ab}-plane of the crystal, indicated in red, while the voltage drop is measured across the two voltage contacts, \textit{V$\,^{+}$} and \textit{V$\,^{-}$}. The \textit{MR(B)} is linear for $\theta=\SI{0}{\degree}$ and tunable to quadratic for $\theta=\SI{90}{\degree}$.}
    	\label{f1}
\end{figure}

We performed low temperature~(\textit{T}), angle-dependent \textit{MR} measurements on single crystalline \ce{TmB4}, which belongs to the rare-earth tetraboride family and crystallizes in a tetragonal structure with space group P4/mbm, 127. The typical layered crystal structure of \ce{TmB4}, with 4 unit cells along the \textit{c}-axis, is shown in fig.~\ref{f1}(a). 
\ce{Tm} atoms lie in the crystalline $ab$-plane, arranged in a Shastry-Sutherland lattice structure\cite{ssl,michi,shin} with approximately equal bond lengths (fig.~\ref{f1}(b)). 
Halfway between the \ce{Tm} layers, planes of boron atoms form a mixture of 4-atom squares and 7-atom rings\cite{shin}. 
There are two different types of boron sites in these planes. 
One type is an exclusive part of the boron plane, whereas the other is part of the boron plane and an octahedral chain along the \textit{c}-axis\cite{shin}. Thus, the crystal structure has both 2D and 3D features.

The low-temperature magnetic measurements carried out earlier\cite{t3,t2,t1,saiprb} on \ce{TmB4} revealed a rich phase diagram with multiple ground states for \textit{B} applied along the \textit{c}-axis.
The ground state is antiferromagnetic~(\textit{AFM}), up to $T=\SI{9.9}{\kelvin}$ (for $B=\SI{0}{\tesla}$) and $B=\SI{1.4}{\tesla}$ (for $T\leq\SI{8}{\kelvin}$).
At higher values of \textit{B} and \textit{T}, the system evolves to various other magnetic ground states, \textit{viz.} a narrow fractional plateau phase (\textit{FPP}), a wide half plateau phase, a modulated phase, and a high-field paramagnetic phase\cite{t3,t2,t1,saiprb}. Recently, specific heat measurements described \textit{FPP}  not as a distinct thermodynamic ground state of \ce{TmB4}, but rather as being degenerate with the \textit{AFM} phase\cite{trinprl}.
Understanding of the various magnetic ground states in \ce{TmB4} has been at the forefront of extensive experimental and theoretical research\cite{t6,t4,t3,t2,prb82,t5,t1,saiprb,chkprb,trinprl}, although transport properties\cite{t2,saiprb,chkprb} are relatively less studied. 
Our previous magnetotransport investigation\cite{saiprb} revealed huge, non-saturating and hysteretic in-plane \textit{MR} (900\% at \SI{7}{\tesla} for \SI{2}{\kelvin}) with signatures of unconventional anomalous Hall effect \cite{saiprb}. The large \textit{MR} along with negative Hall coefficient suggest\cite{saiprb,NbP} that the carriers have high electronic $\mu \sim$\SI{2.9}{\meter^2 \volt^{-1}-\second^{-1}} at \SI{2}{\kelvin}.  

\section{Experiment}
Here, we focus on angle-dependent low-temperature magnetotransport experiments in \ce{TmB4} in its \textit{AFM} phase ($B\le \SI{1.3}{\tesla}$ and $T \le \SI{5}{\kelvin}$). 
A schematic of the experimental arrangement and the main result of this work are shown in fig.~\ref{f1}(c), where $\theta$ is the tilt angle between \textit{B} and \textit{c}-axis. 
We find an unexpected linear \textit{MR}, tunable to quadratic by varying $\theta$.
Single crystals of \ce{TmB4} were grown in a solution growth method using Al solution. Details of the crystal growth can be found elsewhere\cite{t1}. 
For \textit{MR} measurements, the crystal was oriented\cite{saiprb} and cut into pieces with its faces along (001) direction using a tungsten wire. A rectangular piece of dimensions $\sim \SI{0.434}{\milli\meter} \times \SI{0.516}{\milli\meter} \times \SI{0.226}{\milli\meter}$ (weighing $\sim\SI{0.35}{\milli\gram}$) has been used for the measurements.  
The measurement was done in a standard four point probe method using a Quantum Design Physical Property Measurement System (PPMS). The contacts were made with electrically conductive
silver epoxy paste (EpoTeK E4110) and gold wires of diameter $\SI{25}{\micro\meter}$ and $\SI{50}{\micro\meter}$ as connectors
for voltage and current contacts respectively. All measurements were conducted well within the \textit{AFM} phase ($B\le \SI{1.3}{\tesla}$ and $T\le\SI{5}{\kelvin}$).
The angle-dependent magnetotransport measurements were performed by placing the sample on a precision steeper controlled horizontal rotator puck, which can move around an axis perpendicular to \textit{B}. The excitation current (\SI{1.8}{\milli\ampere} and \SI{5.0}{\milli\ampere}) was applied parallel to the \textit{ab}-plane and \textit{B} was applied along various directions, relative to the crystal \textit{c}-axis [see fig.~\ref{f1}(c)]. The linearity of current-voltage was ensured at both \SI{300}{\kelvin} and \SI{2}{\kelvin} prior to the magnetotransport measurements. We found in all cases that the \textit{MR} is minimum at \textit{B} = 0. The raw data of \textit{MR} was then symmetrized to reflect the expected \textit{B} to $-\textit{B}$ invariance, and is plotted in fig.~\ref{f3}(a).
For the anisotropic magnetoresistance (\textit{AMR}) measurements, \textit{R} was measured as the sample was rotated continuously at a fixed \textit{B} and \textit{T}.
 
    \par
    \section{Results and Discussions}
\begin{figure}
    	\includegraphics[width=0.8\linewidth]{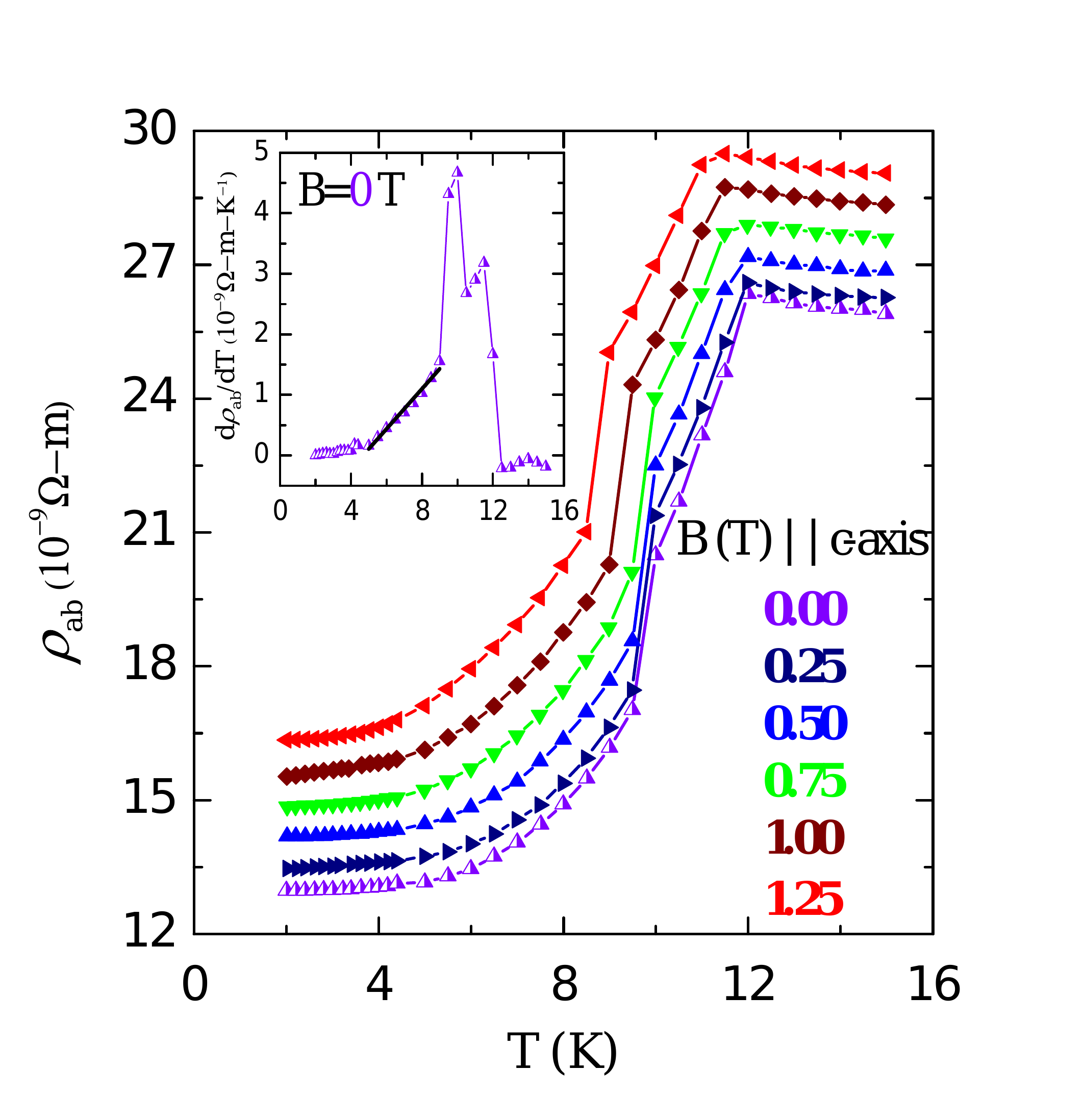}
    	\caption{The temperature and field dependence of electrical resistivity ($\rho$) in the \ce{TmB4} single crystal. Plots of in-plane electrical resistivity ($\rho_\textit{ab}$), measured in various \textit{B}, applied parallel to the crystal \textit{c}-axis, against \textit{T}. At $B=0$, the transition from the paramagnetic state to modulated state occurs at \SI{11.9}{\kelvin} and transition from modulated phase to \textit{AFM} phase occurs at \SI{9.9}{\kelvin}. $\rho_\textit{ab}$ increases and  transition temperature decreases as \textit{B} is increased. For the angle-dependent magnetotransport measurements, we consider the lower part ($2\le \textit{T}\, (\si{\kelvin})\le 5$) of the \twosymb{$\rho_\textit{ab}$}{T} curve. The lines are to guide the eye. Inset: \textit{T}-derivative of $\rho_\textit{ab}(B=0)$ against \textit{T} shows two maxima at the point of inflections of \twosymb{$\rho_\textit{ab}$}{T}, implying the phase transitions. The abscissa of the inset has the same label as the main panel. The black solid line is the linear fit to the experimental data, signifying a \textit{T}\textsuperscript{2} dependence of $\rho_\textit{ab}$, in accordance with the Fermi liquid behavior.  }
    	\label{f2}
\end{figure}

Figure~\ref{f2} depicts the metallic\cite{saiprb} \textit{T} dependence of the in-plane resistivity ($\rho_{ab}$) of \ce{TmB4}
in a longitudinal ($B\parallel c$-axis) field with
varying field strengths.  At room temperature\cite{saiprb}, the zero-field resistivity, $\rho_{ab}(B=0)$ is $\SI{5e-7}{\ohm.\meter}$ and decreases monotonically with decreasing \textit{T} down to \SI{12.9e-9}{\ohm.\meter} at \SI{2}{\kelvin} giving residual resistivity ratio~(RRR=$\frac{\rho_{300 K}}{\rho_{2 K}}$) = $38$. The RRR value is either comparable or even slightly higher than the previously studied \ce{TmB4} crystals\cite{chkprb,t2}, suggesting a good quality crystal with a moderate amount of impurity. At $B=0$, the ratio of $\rho_{ab}$ to the \textit{c}-axis resistivity\cite{saiprb}, $\rho_{c}$, is 0.454 at \SI{2}{\kelvin}.  
Loss of spin-disorder-scattering causes a sudden drop in $\rho_{ab}$ at \SI{11.9}{\kelvin} (at $\textit{B}=0$) as the system undergoes an  magnetic phase transition from the paramagnetic to the modulated phase. Following this second order phase transition, a first order phase transition appears at \SI{9.9}{\kelvin} ($\textit{B}=0$) as the system moves from the magnetically ordered modulated phase to \textit{AFM} state. Under \textit{B}, these transition-\textit{T}s shift to lower values.
As shown in the inset of fig.~\ref{f2}, zero-field $\frac{d\rho_{ab}}{dT}$ shows maxima at $T=\SI{9.9}{\kelvin}$ and \SI{11.9}{\kelvin}, indicative of the above-mentioned phase transitions.
For $T \le \SI{9}{\kelvin}$, $\frac{d\rho_{ab}}{dT}$ decreases linearly with decreasing \textit{T} down to \SI{4}{\kelvin}, implying a $T^2$ variation of resistivity and is almost \textit{T} independent in the lower \textit{T} regime. 
This $T^2$ dependence of resistivity at low \textit{T}, in a metal with magnetic ordering can arise either from $e-e$ scattering or scattering of conduction electrons from magnons \cite{goodings}. A dominant $e-$magnon contribution results in a negative \textit{MR} due to the suppression of magnons\cite{kaul} under \textit{B}. However, unlike magnetic metals, \ce{TmB4} exhibits a positive \textit{MR} and $\rho_{ab}$ increases with \textit{B} (fig.~\ref{f2}). This rules out scattering from magnons as the primary  source of resistivity in \ce{TmB4} and only $e-e$ scattering persists in accordance with Fermi liquid theory ~($\rho_{ab}=\rho_{0}+\beta T^{2}$, where $\rho_{0}$ is the residual resistivity). $\rho_{c}(T)$ also follows a similar $T^2$ behavior\cite{saiprb}. The coefficient 
$\beta$ is inversely proportional to Fermi temperature and is set by the exponent of \textit{T} rather than the residual resistivity\cite{science}. 
While for the in-plane transport, $\beta=\SI{1.6e-10}{\ohm. \meter.\kelvin^{-2}}$, its out-of-plane value is $\SI{83e-10}{\ohm.\meter.\kelvin^{-2}}$. 
    
\begin{figure}
     	\includegraphics[width=0.8\linewidth]{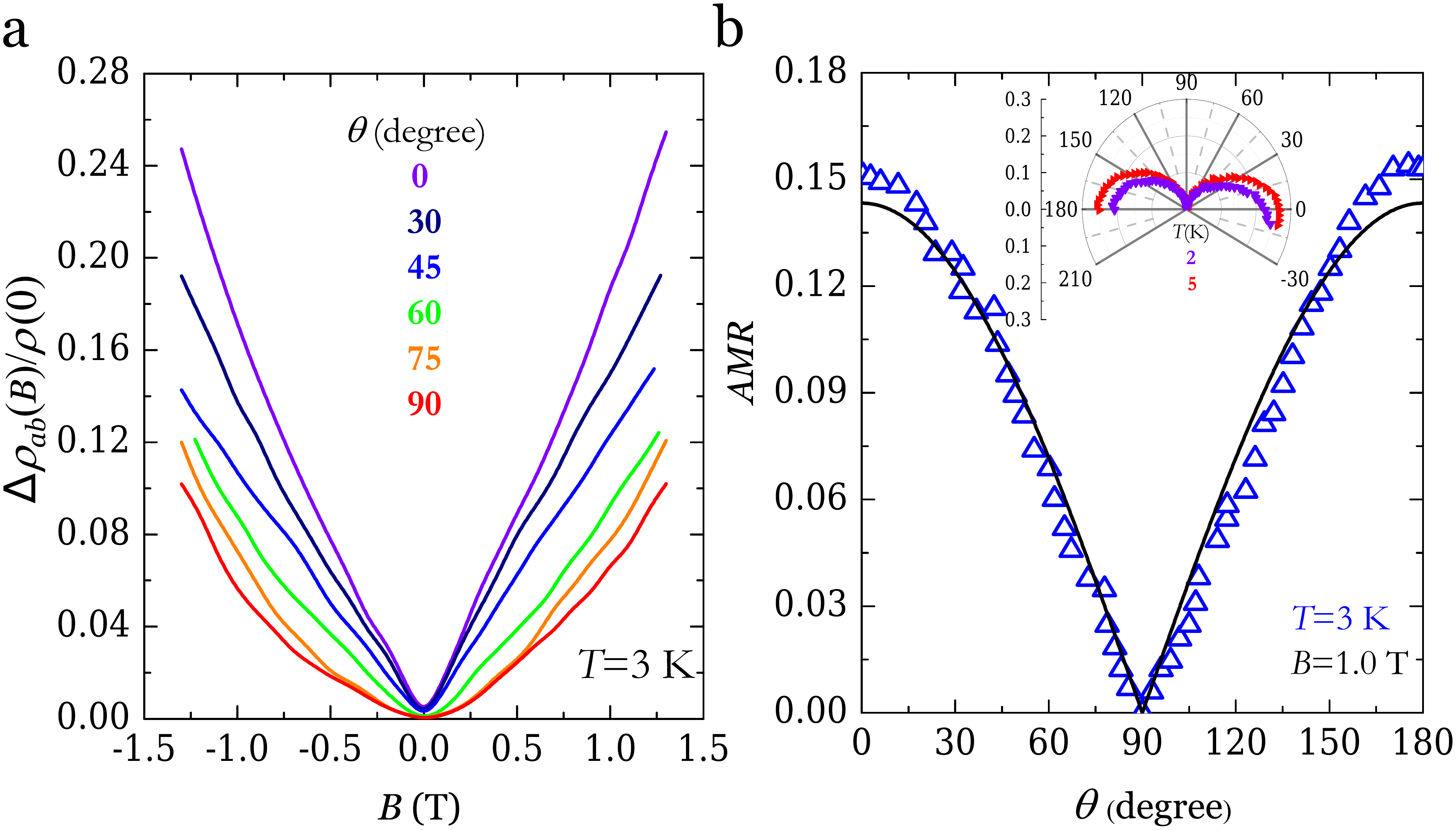}
     	\caption{Angular dependence of \textit{MR}.~(a) A generic \textit{MR}(\textit{B}) isotherm measured at $\textit{T}=\SI{3}{\kelvin}$, under various magnetic field directions. \SI{0}{\degree} (\SI{90}{\degree}) refers to whether \textit{B} is applied parallel (perpendicular) to the crystal's \textit{c}-axis. A linear \textit{MR} can be seen for $\theta= \SI{0}{\degree}$, which gradually moves to a quadratic form for $\theta= \SI{90}{\degree}$. The \textit{MR} is anisotropic. For $\textit{B}=\SI{1.3}{\tesla}$, the \textit{MR} is $\approx$~25{\%} at $\theta= \SI{0}{\degree}$, whereas it is $\approx$~10.3{\%} for $\theta= \SI{90}{\degree}$. (b)~The $\theta$ variation of anisotropic magnetoresistance (\textit{AMR}) (see text) measured at $\textit{B}= \SI{1.0}{\tesla}$ and $\textit{T}= \SI{3}{\kelvin}$. The experimental data can be described by a $\abs{\cos\theta}$ function (solid line) indicating a quasi-2D \textit{FS}\cite{caprb,zhaoprb}. Inset: Polar plot of \textit{AMR}($\theta$) measured at $\textit{T}=  \SI{2}{\kelvin}$ and  $\SI{5}{\kelvin}$, at $\textit{B}=\SI{1.3}{ \tesla}$. The \textit{AMR} shows two-lobes over the full range of $\theta$, suggesting a two-fold symmetry.}
     	\label{f3}
     \end{figure}

Figure~\ref{f3}(a) shows a set of normalized \textit{MR}(\textit{B}) isotherms of \ce{TmB4} with $\theta=\SI{0}{\degree}$ to \SI{90}{\degree}, measured at $T= \SI{3}{\kelvin}$. 
Here, \SI{0}{\degree} (\SI{90}{\degree}) refers to a field \textit{B} applied parallel (perpendicular) to the crystal's \textit{c}-axis (see fig.~\ref{f1}(c)).
Unexpectedly, for $\theta=\SI{0}{\degree}$ to \SI{45}{\degree} the \textit{MR} response is linear all the way down to very small fields. The functional behavior of \textit{MR(B)} changes gradually to quadratic as $\theta\to\SI{90}{\degree}$.  
Whilst the classical \textit{MR} does not have any response when \textit{B} is applied parallel to the excitation current, we observed a close to quadratic growth of \textit{MR} for $B\parallel I\parallel ab$.  The change in \textit{MR} over the \textit{B}-range ($\theta=\SI{90}{\degree}$) is less than 50\% of that observed for $\theta=\SI{0}{\degree}$.
\textit{MR} ($\textit{B}=\SI{1.3}{\tesla}$) is maximum for $\theta=\SI{0}{\degree}$ ($\approx$~25{\%}) and minimum ($\approx$~10.3{\%}) for $\theta=\SI{90}{\degree}$.
\textit{MR}(\textit{B}) essentially shows similar features at other temperatures in the \textit{AFM} phase. 
One of the notable features of the \textit{LMR} in \ce{TmB4} 
is that it persists down to lowest applied field, without showing any signature of crossover to a quadratic behavior with change in \textit{B}, as observed in \ce{CaMnBi2}\cite{caprb} InAs\cite{natmat}, 2DEG\cite{amirag} and CrAs\cite{cras}.  
Instead, this \textit{LMR} is similar to the  super-linear \textit{MR} behavior
observed in non-stoichiometric silver chalcogenides\cite{litnat} , Bi\cite{biSc}, \ce{WTe2}\cite{zhaoprb} and rare-earth diantimonides\cite{redsb}. The slope of $\rho_{ab}(B,\SI{0}{\degree})$ is $\SI{2.21\pm 0.01e-9}{\ohm.\meter.\tesla^{-1}}$ and almost \textit{T}-independent, suggesting the \textit{MR} is not due to the phonon scattering \cite{natmat}.

Furthermore, we find \textit{MR} to be anisotropic.  We define anisotropic magnetoresistance (\textit{AMR}$(\theta)$) as $\frac{R(\theta)-R_{\mathrm{min}}}{R_{\mathrm{min}}}$, where $R(\theta)$ is the resistance at any $\theta$, measured at a constant $B$ and $T$, and $R_{\mathrm{min}}$ is the minimum resistance obtained as $\theta$ is varied.
In fig.~\ref{f3}(b), we show the variation of \textit{AMR}$(\theta)$ at $T= \SI{3}{\kelvin}$ for $B=\SI{1.0}{\tesla}$. \textit{AMR} is maximum for $B\parallel c$-axis and diminishes as \textit{B} is rotated away from the $c$-axis. The data can be satisfactorily fit with a 
 $\abs{\cos\theta}$ dependence. This suggests a (quasi-)$2D$ \textit{FS}\cite{pip,caprb,zhaoprb}, where \textit{MR} responds to the perpendicular component of the applied field, $B~\abs{\cos\theta}$. 
The anisotropic \textit{MR} further suggests an anisotropy in the electronic effective mass\cite{zhaoprb}. \textit{AMR} shows two-fold symmetry (inset fig.~\ref{f3}(b)). 
     \par
\begin{figure}
     	\centering
		\includegraphics[width=0.8\linewidth]{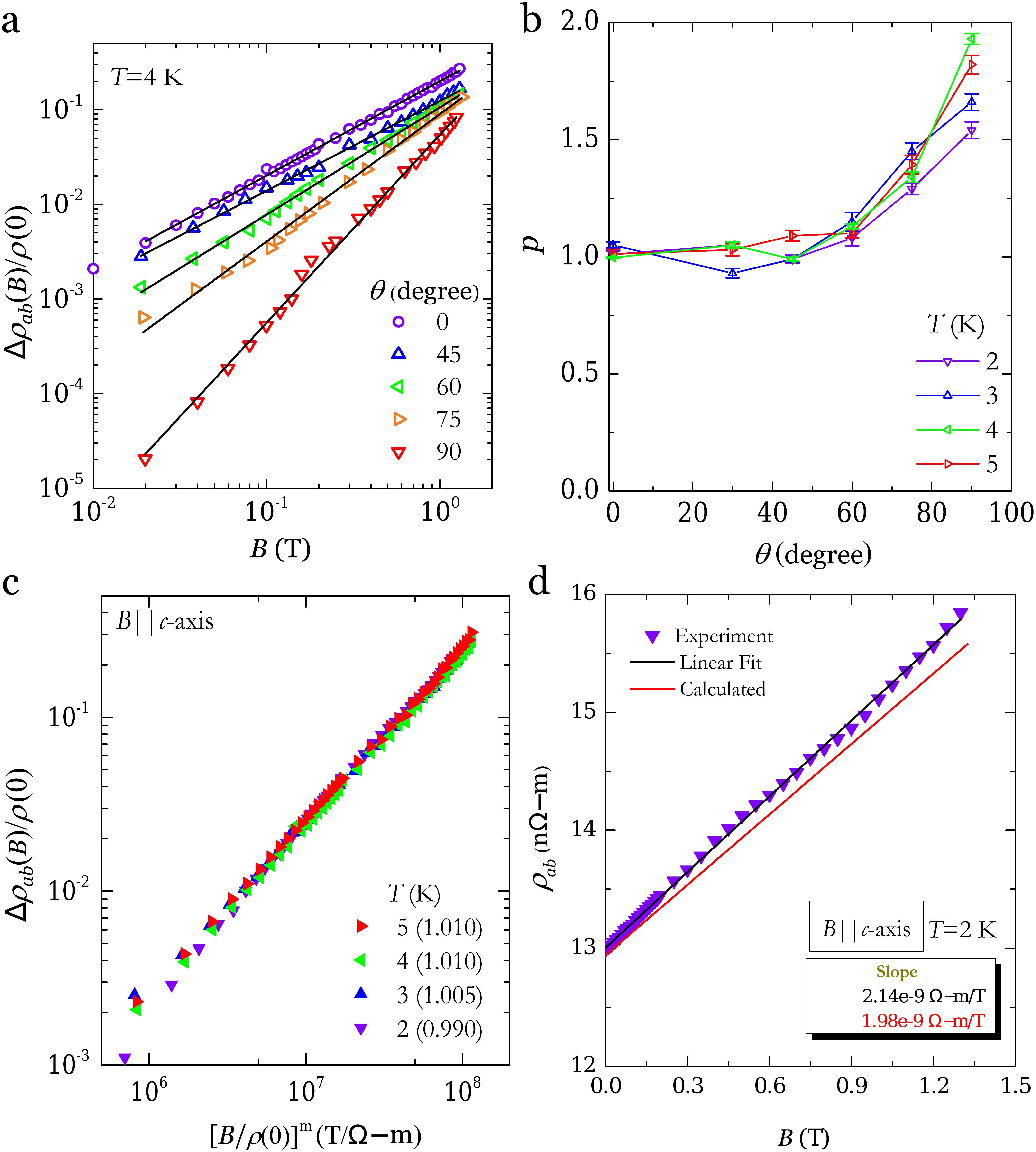}
     	\caption{Analysis of the magnetoresistance data for \ce{TmB4}. (a)~Magnetoresistance isotherm, measured at $\textit{T}=\SI{4}{\kelvin}$, under five different field orientations, shown in a double logarithmic representation. The linearity in the log-log plot suggests a power law behavior, $\textit{MR}= \bigl(\frac{B}{B_{0}} \bigr)^{p}$. The solid lines are the fit of the experimental data to the power law. (b)~The variation of the exponent, \textit{p}, with $\theta$, for various \textit{T}s. For $\theta=\SI{0}{\degree}$, \textit{p} is close to 1, suggesting linear \textit{MR} and gradually moves to a value close to 2, for $\theta=\SI{90}{\degree}$. The error for determining the value of $p$ from the fit (a) is $\sim$ 0.01 and shown in the plot. $p$ varies in a similar manner for all \textit{T}s. The lines are to guide the eye. (c)~\textit{T}-scaling (Kohler rule) of \textit{MR} for \textit{B} along \textit{c}-axis. The values of $m$ used to scale the different \textit{T} \textit{MR} data are mentioned in the parenthesis. (d) Comparison of the  experimental data (violet) and theoretical (red) plot. The experimental data was obtained for $B\parallel c$-axis configuration and measured at $T=\SI{2}{\kelvin}$. The black solid line is the linear fit to the experimental data. The theoretical curve was calculated from Eq.~\ref{eqn:rhoxx} using the values of $N_i$ and $n_e$ (see text). The values of both slopes (inset) agree within 7\%.}
     	\label{f4}
     \end{figure}

To quantify the evolution of \textit{MR} from linear to quadratic, we fit \textit{MR}$(B,\theta)$ to $\bigl(\frac{B}{B_{0}} \bigr)^{p}$. A representative $MR(B)$ plot (in double logarithmic scale), measured at \SI{4}{\kelvin} for different $\theta$ values, is shown in fig.~\ref{f4}(a). 
For $\theta=\SI{0}{\degree}$, $\textit{p} \cong 1$ and gradually grows to $p \approx 2$ (varies between 1.5 to 1.9, for different \textit{T}s) for $\theta=\SI{90}{\degree}$ (fig.~\ref{f4}(b)). Crucially,  $p(\theta)$ varies similarly at all temperatures and has a negligible \textit{T}-dependence within the \textit{AFM} phase (fig.~\ref{f4}(b)).
      
 \textit{MR} (\textit{B}, $\theta=0$) data at different \textit{Ts} can be scaled using the Kohler relation, $\textit{MR}=\alpha(\textit{T})\bigl[\frac{B}{\rho(0)} \bigr]^{m}$ (fig.~\ref{f4}(c)).
The scaling suggests that the carriers with single salient relaxation time\cite{pip} govern magnetotransport for $\textit{B}\parallel\textit{c}$-axis in the \textit{AFM} phase.
Furthermore, this robust \textit{T}-scaling, using a single $\alpha$, adds credence to the relative \textit{T}-insensitivity of \textit{LMR} and implies negligible phononic contributions.
Therefore, the measured \textit{MR} is primarily governed by scattering of conduction electrons by impurities. 
\par

\begin{figure}[htbp]
 \centering
\includegraphics[width=0.7\linewidth]{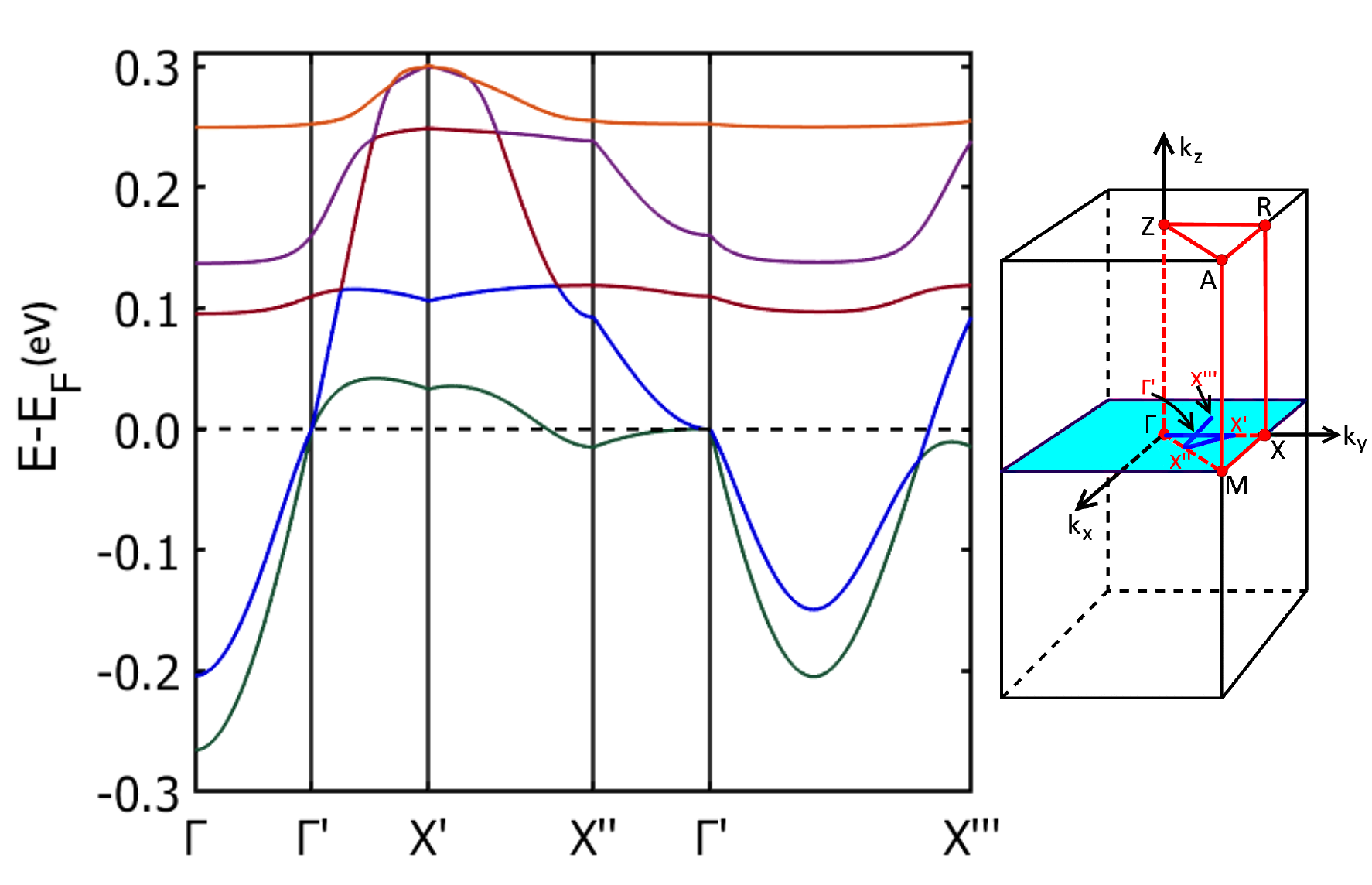}
\caption{Band structure for \ce{TmB4} from the spin polarized DFT calculation using GGA+U method, shows dispersion in the \textit{ab} plane. The right panel shows the path (in blue) taken in the \textit{BZ}, $\Gamma$--$\Gamma'$--X$'$--X$''$--$\Gamma'$--X$'''$. A band crossing is observed exactly at $E_F$ with linear dispersion in the $k_x-k_y$ plane at $\Gamma'$. }
\label{f5}
\end{figure}

The origin of \textit{LMR} in \ce{TmB4} is not entirely clear, but it is plausible that  Abrikosov's theory of quantum linear \textit{MR}\cite{ab1,ab2,ab3,ab4} can be invoked for this purpose, considering the topology of the Fermi surface(\textit{FS}) of \ce{TmB4} \cite{shin}.
The presence of two symmetry related small pockets (as evident from the spin-polarized DFT calculation using GGA+U method\footnote{For calculation details please see Ref.~\citenum{shin}}) in the
$k_x-k_y$ plane of the Brillouin Zone (\textit{BZ}) along the $\Gamma-X$ direction (labeled $\Gamma'$ in fig.~\ref{f5}),
with an approximately linear crossing of two bands
at the Fermi energy, $E_F$ (within the numerical accuracy) (Fig.~\ref{f5}) is of particular interest here. The low density and small effective mass of the carriers due to the linear band crossing, ensure that they can be confined to the lowest Landau level, and thus reach the extreme quantum limit even at small (longitudinal) applied fields. This results to a LMR \cite{ab1,ab2,ab3,ab4} and is given by,

\begin{equation}
\rho_{xx}=\frac{N_iB}{\pi cen_e^2}
\label{eqn:rhoxx}
\end{equation}
provided that the carrier concentration ($n_e$) satisfies
$n_e \lesssim \left( \frac{m_{zz}}{m_{xy}}\right)^{\frac{1}{2}} \left( \frac{eB}{\hbar c}\right)^{\frac{3}{2}}$,
$m_{zz}$ and $m_{xy}$ are the effective mass of
the carriers for motion along $k_z$ and in the $k_x-k_y$ plane, respectively and $N_i$ ($\ll n_e$) is the density of static scattering centers.

The low effective mass of the carriers further gives a \textit{T}-limit for lowest Landau level confinement (see Supplementary information) which is indeed satisfied in our experiments\footnote{Above this \textit{T}-value, we previously observed a quadratic \textit{MR} for $B\parallel c$-axis. Please see ~Ref.\citenum{saiprb} for details}.
At small fields, due to the low effective mass of the electrons from the Fermi pockets, and  consequently their high cyclotron frequency, the linear contribution dominates over the usual quadratic \textit{MR} from the rest of the \textit{FS}\cite{ab4}.  
Using the values of carrier density and their effective masses estimated from band structure calculations,\footnote{Estimating $n_e^{\text{pocket}}$ is tricky, since
the pocket is reduced to a point in the zero-field first principle calculations. However, as a rough
estimate, we can use the calculated value for the other smallest Fermi pockets\cite{shin}, viz.,
$n_e ~(\sim\SI{e24}{\meter^{-3}}$.) } as well as impurity concentrations\footnote{A comparison between the residual resistivity (\textit{RR}) of our sample and similarly grown \ce{TmB4} crystals with known impurity density\cite{jjap} allows us to estimate $N_i$ (assuming the \textit{RR} arises solely from scattering of electrons off impurity), as $\SI{e21}{\meter^{-3}}$. For details, see Supplementary information} from sample preparation conditions, Eq.~\ref{eqn:rhoxx} yields an \textit{MR(B)} that is in agreement with the experimentally observed magnetotransport data in \ce{TmB4} (fig.~\ref{f4}(d)). The compliance of \textit{MR (B, $\theta=0$)} to Kohler scaling provides further support to the assumption that magnetotransport at small longitudinal fields is dominated by charge carriers from identical Fermi pockets.

The mechanism identified above explains another intriguing feature of  \ce{TmB4} -- the absence of  Shubnikov-de Haas (SdH) oscillations in the observed {\it MR} data. Since the extreme quantum limit is already reached at very small fields for the pocket under consideration, there are no Landau level crossings of the \textit{FS} with increasing field, and consequently no SdH oscillations. In principle, the SdH oscillations should be observed for \textit{B} along \textit{ab}-plane, but we could not reach the required \textit{B}, due to strong magnetic fluctuations and experimental limitations. 

Finally, the absence of LMR for transverse magnetic fields can also be understood from the anisotropic \textit{FS} topology. Being a layered material, the small pockets in the $k_x-k_y$ plane of \ce{TmB4} are believed to originate from the overlap of bands close to the \textit{FS} due to the inter-layer coupling. Consequently, there are no such pockets at corresponding points on the surface of the \textit{BZ} in the \textit{XY} plane.
Since magnetotransport of a solid is governed by the external cross section of the \textit{FS} along the field direction\cite{caprb}, only the quadratic contribution of the total conductivity persists for \textit{B} applied along the principal plane.
This picture, based on the topology of the \textit{FS} of \ce{TmB4}, qualitatively explains the experimental observation 
of tuning \textit{MR} from quadratic to  linear as the field direction is rotated. 

It should be noted that  the above discussion is a plausible, rather than a rigorous elucidation for the origin of \textit{LMR} in \ce{TmB4}. The present
explanation depends crucially on the existence of a linear band crossing very close to the \textit{FS} in the $k_x-k_y$ plane. Unfortunately, DFT is unable to capture the effects of strong correlations with high accuracy, therefore one must regard the interpretation as tentative and a much rigorous analytic calculation is indeed required for better insight into the problem. However, it is interesting that  the present approach based on anisotropic \textit{FS} topology within the quantum linear magnetoresistance framework is consistent with the experimental observations. It thus provides a useful platform for further studies of this compelling phenomenon. 
\section{Summary}
In summary, we have discussed the tuning of \textit{MR} from linear to quadratic in single crystalline metal, \ce{TmB4}, by rotating \textit{B} relative to the crystal \textit{c}-axis. 
We give a plausible explanation of the \textit{LMR} in this metallic system based on its \textit{FS} topology within the quantum linear magnetoresistance picture, which predominantly holds true for semimetals and topological insulators. We argued that the linear dispersion near $E_{F}$ and the subsequent Fermi pocket in the \textit{FS} of \ce{TmB4}, arising from its layered structure, give rise to a \textit{LMR} in an otherwise normal metal and its complex \textit{FS} topology governs the tuning of in-plane \textit{MR} from quadratic to linear.
  
\section*{acknowledgements}
The work in Singapore is supported by a grant (MOE2014-T2-2-112) from Ministry of Education,
Singapore and the National Research Foundation (NRF), NRF-Investigatorship (NRFNRFI2015-04). SM and JGSK acknowledge stimulating discussions with Arthur Ramirez, Alexander Petrovic, Xian Yang Tee, Jennifer Trinh and Bhartendu Satywali. The work at UCSC was supported by the U.S. Department of Energy (BES) under Award number DE-FG02-06ER46319. Work performed at Ames Laboratory was supported by the U.S. Department of Energy, Office of Basic Energy Science, Division of Materials Sciences and Engineering. Ames Laboratory is operated for the U.S. Department of Energy by Iowa State University under Contract No. DE-AC02-07CH11358.


\begin{thebibliography}{52}%
\makeatletter
\providecommand \@ifxundefined [1]{%
 \@ifx{#1\undefined}
}%
\providecommand \@ifnum [1]{%
 \ifnum #1\expandafter \@firstoftwo
 \else \expandafter \@secondoftwo
 \fi
}%
\providecommand \@ifx [1]{%
 \ifx #1\expandafter \@firstoftwo
 \else \expandafter \@secondoftwo
 \fi
}%
\providecommand \natexlab [1]{#1}%
\providecommand \enquote  [1]{``#1''}%
\providecommand \bibnamefont  [1]{#1}%
\providecommand \bibfnamefont [1]{#1}%
\providecommand \citenamefont [1]{#1}%
\providecommand \href@noop [0]{\@secondoftwo}%
\providecommand \href [0]{\begingroup \@sanitize@url \@href}%
\providecommand \@href[1]{\@@startlink{#1}\@@href}%
\providecommand \@@href[1]{\endgroup#1\@@endlink}%
\providecommand \@sanitize@url [0]{\catcode `\\12\catcode `\$12\catcode
  `\&12\catcode `\#12\catcode `\^12\catcode `\_12\catcode `\%12\relax}%
\providecommand \@@startlink[1]{}%
\providecommand \@@endlink[0]{}%
\providecommand \url  [0]{\begingroup\@sanitize@url \@url }%
\providecommand \@url [1]{\endgroup\@href {#1}{\urlprefix }}%
\providecommand \urlprefix  [0]{URL }%
\providecommand \Eprint [0]{\href }%
\providecommand \doibase [0]{http://dx.doi.org/}%
\providecommand \selectlanguage [0]{\@gobble}%
\providecommand \bibinfo  [0]{\@secondoftwo}%
\providecommand \bibfield  [0]{\@secondoftwo}%
\providecommand \translation [1]{[#1]}%
\providecommand \BibitemOpen [0]{}%
\providecommand \bibitemStop [0]{}%
\providecommand \bibitemNoStop [0]{.\EOS\space}%
\providecommand \EOS [0]{\spacefactor3000\relax}%
\providecommand \BibitemShut  [1]{\csname bibitem#1\endcsname}%
\let\auto@bib@innerbib\@empty
\bibitem [{\citenamefont {Niu}\ \emph {et~al.}(2017)\citenamefont {Niu},
  \citenamefont {Yu}, \citenamefont {Yip}, \citenamefont {Lim}, \citenamefont
  {Kotegawa}, \citenamefont {Matsuoka}, \citenamefont {Sugawara}, \citenamefont
  {Tou}, \citenamefont {Yanase},\ and\ \citenamefont {Goh}}]{cras}%
  \BibitemOpen
  \bibfield  {author} {\bibinfo {author} {\bibfnamefont {Q.}~\bibnamefont
  {Niu}}, \bibinfo {author} {\bibfnamefont {W.~C.}\ \bibnamefont {Yu}},
  \bibinfo {author} {\bibfnamefont {K.~Y.}\ \bibnamefont {Yip}}, \bibinfo
  {author} {\bibfnamefont {Z.~L.}\ \bibnamefont {Lim}}, \bibinfo {author}
  {\bibfnamefont {H.}~\bibnamefont {Kotegawa}}, \bibinfo {author}
  {\bibfnamefont {E.}~\bibnamefont {Matsuoka}}, \bibinfo {author}
  {\bibfnamefont {H.}~\bibnamefont {Sugawara}}, \bibinfo {author}
  {\bibfnamefont {H.}~\bibnamefont {Tou}}, \bibinfo {author} {\bibfnamefont
  {Y.}~\bibnamefont {Yanase}}, \ and\ \bibinfo {author} {\bibfnamefont {S.~K.}\
  \bibnamefont {Goh}},\ }\href {\doibase 10.1038/ncomms15358} {\bibfield
  {journal} {\bibinfo  {journal} {Nat. Commun.}\ }\textbf {\bibinfo {volume}
  {8}},\ \bibinfo {pages} {15358} (\bibinfo {year} {2017})}\BibitemShut
  {NoStop}%
\bibitem [{\citenamefont {Pippard}(1989)}]{pip}%
  \BibitemOpen
  \bibfield  {author} {\bibinfo {author} {\bibfnamefont {A.~B.}\ \bibnamefont
  {Pippard}},\ }\href@noop {} {\emph {\bibinfo {title} {Magnetoresistance in
  {Metals}}}}\ (\bibinfo  {publisher} {Cambridge University Press},\ \bibinfo
  {address} {Cambridge, USA},\ \bibinfo {year} {1989})\ p.\ \bibinfo {pages}
  {253}\BibitemShut {NoStop}%
\bibitem [{\citenamefont {Parish}\ and\ \citenamefont
  {Littlewood}(2003)}]{parnat}%
  \BibitemOpen
  \bibfield  {author} {\bibinfo {author} {\bibfnamefont {M.~M.}\ \bibnamefont
  {Parish}}\ and\ \bibinfo {author} {\bibfnamefont {P.~B.}\ \bibnamefont
  {Littlewood}},\ }\href@noop {} {\bibfield  {journal} {\bibinfo  {journal}
  {Nature}\ }\textbf {\bibinfo {volume} {426}},\ \bibinfo {pages} {162}
  (\bibinfo {year} {2003})}\BibitemShut {NoStop}%
\bibitem [{\citenamefont {Parish}\ and\ \citenamefont
  {Littlewood}(2005)}]{parprb}%
  \BibitemOpen
  \bibfield  {author} {\bibinfo {author} {\bibfnamefont {M.~M.}\ \bibnamefont
  {Parish}}\ and\ \bibinfo {author} {\bibfnamefont {P.~B.}\ \bibnamefont
  {Littlewood}},\ }\href@noop {} {\bibfield  {journal} {\bibinfo  {journal}
  {Phys. Rev. B}\ }\textbf {\bibinfo {volume} {72}},\ \bibinfo {pages} {094417}
  (\bibinfo {year} {2005})}\BibitemShut {NoStop}%
\bibitem [{\citenamefont {Yan}\ \emph {et~al.}(2013)\citenamefont {Yan},
  \citenamefont {Wang}, \citenamefont {Yu},\ and\ \citenamefont {Liao}}]{apl}%
  \BibitemOpen
  \bibfield  {author} {\bibinfo {author} {\bibfnamefont {Y.}~\bibnamefont
  {Yan}}, \bibinfo {author} {\bibfnamefont {L.-X.}\ \bibnamefont {Wang}},
  \bibinfo {author} {\bibfnamefont {D.-P.}\ \bibnamefont {Yu}}, \ and\ \bibinfo
  {author} {\bibfnamefont {Z.-M.}\ \bibnamefont {Liao}},\ }\href@noop {}
  {\bibfield  {journal} {\bibinfo  {journal} {Appl. Phys. Lett.}\ }\textbf
  {\bibinfo {volume} {103}},\ \bibinfo {pages} {033106} (\bibinfo {year}
  {2013})}\BibitemShut {NoStop}%
\bibitem [{\citenamefont {Alekseev}\ \emph {et~al.}(2015)\citenamefont
  {Alekseev}, \citenamefont {Dmitriev}, \citenamefont {Gornyi}, \citenamefont
  {Kachorovskii}, \citenamefont {Narozhny}, \citenamefont {Sch\"utt},\ and\
  \citenamefont {Titov}}]{prl2015}%
  \BibitemOpen
  \bibfield  {author} {\bibinfo {author} {\bibfnamefont {P.~S.}\ \bibnamefont
  {Alekseev}}, \bibinfo {author} {\bibfnamefont {A.~P.}\ \bibnamefont
  {Dmitriev}}, \bibinfo {author} {\bibfnamefont {I.~V.}\ \bibnamefont
  {Gornyi}}, \bibinfo {author} {\bibfnamefont {V.~Y.}\ \bibnamefont
  {Kachorovskii}}, \bibinfo {author} {\bibfnamefont {B.~N.}\ \bibnamefont
  {Narozhny}}, \bibinfo {author} {\bibfnamefont {M.}~\bibnamefont {Sch\"utt}},
  \ and\ \bibinfo {author} {\bibfnamefont {M.}~\bibnamefont {Titov}},\
  }\href@noop {} {\bibfield  {journal} {\bibinfo  {journal} {Phys. Rev. Lett.}\
  }\textbf {\bibinfo {volume} {114}},\ \bibinfo {pages} {156601} (\bibinfo
  {year} {2015})}\BibitemShut {NoStop}%
\bibitem [{\citenamefont {Narayanan}\ \emph {et~al.}(2015)\citenamefont
  {Narayanan}, \citenamefont {Watson}, \citenamefont {Blake}, \citenamefont
  {Bruyant}, \citenamefont {Drigo}, \citenamefont {Chen}, \citenamefont
  {Prabhakaran}, \citenamefont {Yan}, \citenamefont {Felser}, \citenamefont
  {Kong}, \citenamefont {Canfield},\ and\ \citenamefont {Coldea}}]{dsm2}%
  \BibitemOpen
  \bibfield  {author} {\bibinfo {author} {\bibfnamefont {A.}~\bibnamefont
  {Narayanan}}, \bibinfo {author} {\bibfnamefont {M.~D.}\ \bibnamefont
  {Watson}}, \bibinfo {author} {\bibfnamefont {S.~F.}\ \bibnamefont {Blake}},
  \bibinfo {author} {\bibfnamefont {N.}~\bibnamefont {Bruyant}}, \bibinfo
  {author} {\bibfnamefont {L.}~\bibnamefont {Drigo}}, \bibinfo {author}
  {\bibfnamefont {Y.~L.}\ \bibnamefont {Chen}}, \bibinfo {author}
  {\bibfnamefont {D.}~\bibnamefont {Prabhakaran}}, \bibinfo {author}
  {\bibfnamefont {B.}~\bibnamefont {Yan}}, \bibinfo {author} {\bibfnamefont
  {C.}~\bibnamefont {Felser}}, \bibinfo {author} {\bibfnamefont
  {T.}~\bibnamefont {Kong}}, \bibinfo {author} {\bibfnamefont {P.~C.}\
  \bibnamefont {Canfield}}, \ and\ \bibinfo {author} {\bibfnamefont {A.~I.}\
  \bibnamefont {Coldea}},\ }\href {\doibase 10.1103/PhysRevLett.114.117201}
  {\bibfield  {journal} {\bibinfo  {journal} {Phys. Rev. Lett.}\ }\textbf
  {\bibinfo {volume} {114}},\ \bibinfo {pages} {117201} (\bibinfo {year}
  {2015})}\BibitemShut {NoStop}%
\bibitem [{\citenamefont {{Song}}\ \emph {et~al.}(2015)\citenamefont {{Song}},
  \citenamefont {{Refael}},\ and\ \citenamefont {{Lee}}}]{justin}%
  \BibitemOpen
  \bibfield  {author} {\bibinfo {author} {\bibfnamefont {J.~C.~W.}\
  \bibnamefont {{Song}}}, \bibinfo {author} {\bibfnamefont {G.}~\bibnamefont
  {{Refael}}}, \ and\ \bibinfo {author} {\bibfnamefont {P.~A.}\ \bibnamefont
  {{Lee}}},\ }\href@noop {} {\bibfield  {journal} {\bibinfo  {journal} {Phys.
  Rev. B}\ }\textbf {\bibinfo {volume} {92}},\ \bibinfo {pages} {180204}
  (\bibinfo {year} {2015})}\BibitemShut {NoStop}%
\bibitem [{\citenamefont {Kisslinger}\ \emph {et~al.}(2017)\citenamefont
  {Kisslinger}, \citenamefont {Ott},\ and\ \citenamefont {Weber}}]{prb95}%
  \BibitemOpen
  \bibfield  {author} {\bibinfo {author} {\bibfnamefont {F.}~\bibnamefont
  {Kisslinger}}, \bibinfo {author} {\bibfnamefont {C.}~\bibnamefont {Ott}}, \
  and\ \bibinfo {author} {\bibfnamefont {H.~B.}\ \bibnamefont {Weber}},\ }\href
  {\doibase 10.1103/PhysRevB.95.024204} {\bibfield  {journal} {\bibinfo
  {journal} {Phys. Rev. B}\ }\textbf {\bibinfo {volume} {95}},\ \bibinfo
  {pages} {024204} (\bibinfo {year} {2017})}\BibitemShut {NoStop}%
\bibitem [{\citenamefont {Khouri}\ \emph {et~al.}(2016)\citenamefont {Khouri},
  \citenamefont {Zeitler}, \citenamefont {Reichl}, \citenamefont {Wegscheider},
  \citenamefont {Hussey}, \citenamefont {Wiedmann},\ and\ \citenamefont
  {Maan}}]{prl117}%
  \BibitemOpen
  \bibfield  {author} {\bibinfo {author} {\bibfnamefont {T.}~\bibnamefont
  {Khouri}}, \bibinfo {author} {\bibfnamefont {U.}~\bibnamefont {Zeitler}},
  \bibinfo {author} {\bibfnamefont {C.}~\bibnamefont {Reichl}}, \bibinfo
  {author} {\bibfnamefont {W.}~\bibnamefont {Wegscheider}}, \bibinfo {author}
  {\bibfnamefont {N.~E.}\ \bibnamefont {Hussey}}, \bibinfo {author}
  {\bibfnamefont {S.}~\bibnamefont {Wiedmann}}, \ and\ \bibinfo {author}
  {\bibfnamefont {J.~C.}\ \bibnamefont {Maan}},\ }\href@noop {} {\bibfield
  {journal} {\bibinfo  {journal} {Phys. Rev. Lett.}\ }\textbf {\bibinfo
  {volume} {117}},\ \bibinfo {pages} {256601} (\bibinfo {year}
  {2016})}\BibitemShut {NoStop}%
\bibitem [{\citenamefont {Abrikosov}(1969)}]{ab1}%
  \BibitemOpen
  \bibfield  {author} {\bibinfo {author} {\bibfnamefont {A.~A.}\ \bibnamefont
  {Abrikosov}},\ }\href@noop {} {\bibfield  {journal} {\bibinfo  {journal}
  {Sov. Phys. JETP}\ }\textbf {\bibinfo {volume} {29}},\ \bibinfo {pages} {746}
  (\bibinfo {year} {1969})}\BibitemShut {NoStop}%
\bibitem [{\citenamefont {Abrikosov}(1998)}]{ab2}%
  \BibitemOpen
  \bibfield  {author} {\bibinfo {author} {\bibfnamefont {A.~A.}\ \bibnamefont
  {Abrikosov}},\ }\href@noop {} {\bibfield  {journal} {\bibinfo  {journal}
  {Phys. Rev. B}\ }\textbf {\bibinfo {volume} {58}},\ \bibinfo {pages} {2788}
  (\bibinfo {year} {1998})}\BibitemShut {NoStop}%
\bibitem [{\citenamefont {Liang}\ \emph {et~al.}(2015)\citenamefont {Liang},
  \citenamefont {Gibson}, \citenamefont {Ali}, \citenamefont {Liu},
  \citenamefont {Cava},\ and\ \citenamefont {Ong}}]{dsm}%
  \BibitemOpen
  \bibfield  {author} {\bibinfo {author} {\bibfnamefont {T.}~\bibnamefont
  {Liang}}, \bibinfo {author} {\bibfnamefont {Q.}~\bibnamefont {Gibson}},
  \bibinfo {author} {\bibfnamefont {M.~N.}\ \bibnamefont {Ali}}, \bibinfo
  {author} {\bibfnamefont {M.}~\bibnamefont {Liu}}, \bibinfo {author}
  {\bibfnamefont {R.~J.}\ \bibnamefont {Cava}}, \ and\ \bibinfo {author}
  {\bibfnamefont {N.~P.}\ \bibnamefont {Ong}},\ }\href@noop {} {\bibfield
  {journal} {\bibinfo  {journal} {Nat. Mater.}\ }\textbf {\bibinfo {volume}
  {14}},\ \bibinfo {pages} {280} (\bibinfo {year} {2015})}\BibitemShut
  {NoStop}%
\bibitem [{\citenamefont {Yang}\ \emph {et~al.}(1999)\citenamefont {Yang},
  \citenamefont {Liu}, \citenamefont {Hong}, \citenamefont {Reich},
  \citenamefont {Searson},\ and\ \citenamefont {Chien}}]{biSc}%
  \BibitemOpen
  \bibfield  {author} {\bibinfo {author} {\bibfnamefont {F.~Y.}\ \bibnamefont
  {Yang}}, \bibinfo {author} {\bibfnamefont {K.}~\bibnamefont {Liu}}, \bibinfo
  {author} {\bibfnamefont {K.}~\bibnamefont {Hong}}, \bibinfo {author}
  {\bibfnamefont {D.~H.}\ \bibnamefont {Reich}}, \bibinfo {author}
  {\bibfnamefont {P.~C.}\ \bibnamefont {Searson}}, \ and\ \bibinfo {author}
  {\bibfnamefont {C.~L.}\ \bibnamefont {Chien}},\ }\href@noop {} {\bibfield
  {journal} {\bibinfo  {journal} {Science}\ }\textbf {\bibinfo {volume}
  {284}},\ \bibinfo {pages} {1335} (\bibinfo {year} {1999})}\BibitemShut
  {NoStop}%
\bibitem [{\citenamefont {Xu}\ \emph {et~al.}(1997)\citenamefont {Xu},
  \citenamefont {Husmann}, \citenamefont {Rosenbaum}, \citenamefont {Saboungi},
  \citenamefont {Enderby},\ and\ \citenamefont {Littlewood}}]{litnat}%
  \BibitemOpen
  \bibfield  {author} {\bibinfo {author} {\bibfnamefont {R.}~\bibnamefont
  {Xu}}, \bibinfo {author} {\bibfnamefont {A.}~\bibnamefont {Husmann}},
  \bibinfo {author} {\bibfnamefont {T.~F.}\ \bibnamefont {Rosenbaum}}, \bibinfo
  {author} {\bibfnamefont {M.-L.}\ \bibnamefont {Saboungi}}, \bibinfo {author}
  {\bibfnamefont {J.~E.}\ \bibnamefont {Enderby}}, \ and\ \bibinfo {author}
  {\bibfnamefont {P.~B.}\ \bibnamefont {Littlewood}},\ }\href@noop {}
  {\bibfield  {journal} {\bibinfo  {journal} {Nature}\ }\textbf {\bibinfo
  {volume} {390}},\ \bibinfo {pages} {57} (\bibinfo {year} {1997})}\BibitemShut
  {NoStop}%
\bibitem [{\citenamefont {Kopelevich}\ \emph {et~al.}(2013)\citenamefont
  {Kopelevich}, \citenamefont {da~Silva}, \citenamefont {Camargo},\ and\
  \citenamefont {Alexandrov}}]{graphi}%
  \BibitemOpen
  \bibfield  {author} {\bibinfo {author} {\bibfnamefont {Y.}~\bibnamefont
  {Kopelevich}}, \bibinfo {author} {\bibfnamefont {R.~R.}\ \bibnamefont
  {da~Silva}}, \bibinfo {author} {\bibfnamefont {B.~C.}\ \bibnamefont
  {Camargo}}, \ and\ \bibinfo {author} {\bibfnamefont {A.~S.}\ \bibnamefont
  {Alexandrov}},\ }\href@noop {} {\bibfield  {journal} {\bibinfo  {journal} {J.
  Phys. Condens. Matter}\ }\textbf {\bibinfo {volume} {25}},\ \bibinfo {pages}
  {466004} (\bibinfo {year} {2013})}\BibitemShut {NoStop}%
\bibitem [{\citenamefont {Friedman}\ \emph {et~al.}(2010)\citenamefont
  {Friedman}, \citenamefont {Tedesco}, \citenamefont {Campbell}, \citenamefont
  {Culbertson}, \citenamefont {Aifer}, \citenamefont {Perkins}, \citenamefont
  {Myers-Ward}, \citenamefont {Hite}, \citenamefont {Eddy}, \citenamefont
  {Jernigan},\ and\ \citenamefont {Gaskill}}]{graNL}%
  \BibitemOpen
  \bibfield  {author} {\bibinfo {author} {\bibfnamefont {A.~L.}\ \bibnamefont
  {Friedman}}, \bibinfo {author} {\bibfnamefont {J.~L.}\ \bibnamefont
  {Tedesco}}, \bibinfo {author} {\bibfnamefont {P.~M.}\ \bibnamefont
  {Campbell}}, \bibinfo {author} {\bibfnamefont {J.~C.}\ \bibnamefont
  {Culbertson}}, \bibinfo {author} {\bibfnamefont {E.}~\bibnamefont {Aifer}},
  \bibinfo {author} {\bibfnamefont {F.~K.}\ \bibnamefont {Perkins}}, \bibinfo
  {author} {\bibfnamefont {R.~L.}\ \bibnamefont {Myers-Ward}}, \bibinfo
  {author} {\bibfnamefont {J.~K.}\ \bibnamefont {Hite}}, \bibinfo {author}
  {\bibfnamefont {C.~R.}\ \bibnamefont {Eddy}}, \bibinfo {author}
  {\bibfnamefont {G.~G.}\ \bibnamefont {Jernigan}}, \ and\ \bibinfo {author}
  {\bibfnamefont {D.~K.}\ \bibnamefont {Gaskill}},\ }\href@noop {} {\bibfield
  {journal} {\bibinfo  {journal} {Nano Lett.}\ }\textbf {\bibinfo {volume}
  {10}},\ \bibinfo {pages} {3962} (\bibinfo {year} {2010})}\BibitemShut
  {NoStop}%
\bibitem [{\citenamefont {Wang}\ \emph
  {et~al.}(2012{\natexlab{a}})\citenamefont {Wang}, \citenamefont {Graf},
  \citenamefont {Wang}, \citenamefont {Lei}, \citenamefont {Tozer},\ and\
  \citenamefont {Petrovic}}]{caprb}%
  \BibitemOpen
  \bibfield  {author} {\bibinfo {author} {\bibfnamefont {K.}~\bibnamefont
  {Wang}}, \bibinfo {author} {\bibfnamefont {D.}~\bibnamefont {Graf}}, \bibinfo
  {author} {\bibfnamefont {L.}~\bibnamefont {Wang}}, \bibinfo {author}
  {\bibfnamefont {H.}~\bibnamefont {Lei}}, \bibinfo {author} {\bibfnamefont
  {S.~W.}\ \bibnamefont {Tozer}}, \ and\ \bibinfo {author} {\bibfnamefont
  {C.}~\bibnamefont {Petrovic}},\ }\href@noop {} {\bibfield  {journal}
  {\bibinfo  {journal} {Phys. Rev. B}\ }\textbf {\bibinfo {volume} {85}},\
  \bibinfo {pages} {041101} (\bibinfo {year} {2012}{\natexlab{a}})}\BibitemShut
  {NoStop}%
\bibitem [{\citenamefont {Delmo}\ \emph {et~al.}(2009)\citenamefont {Delmo},
  \citenamefont {Yamamoto}, \citenamefont {Kasai}, \citenamefont {Ono},\ and\
  \citenamefont {Kobayashi}}]{sinat}%
  \BibitemOpen
  \bibfield  {author} {\bibinfo {author} {\bibfnamefont {M.~P.}\ \bibnamefont
  {Delmo}}, \bibinfo {author} {\bibfnamefont {S.}~\bibnamefont {Yamamoto}},
  \bibinfo {author} {\bibfnamefont {S.}~\bibnamefont {Kasai}}, \bibinfo
  {author} {\bibfnamefont {T.}~\bibnamefont {Ono}}, \ and\ \bibinfo {author}
  {\bibfnamefont {K.}~\bibnamefont {Kobayashi}},\ }\href@noop {} {\bibfield
  {journal} {\bibinfo  {journal} {Nature}\ }\textbf {\bibinfo {volume} {457}},\
  \bibinfo {pages} {1112} (\bibinfo {year} {2009})}\BibitemShut {NoStop}%
\bibitem [{\citenamefont {Hu}\ and\ \citenamefont {Rosenbaum}(2008)}]{natmat}%
  \BibitemOpen
  \bibfield  {author} {\bibinfo {author} {\bibfnamefont {J.}~\bibnamefont
  {Hu}}\ and\ \bibinfo {author} {\bibfnamefont {T.~F.}\ \bibnamefont
  {Rosenbaum}},\ }\href@noop {} {\bibfield  {journal} {\bibinfo  {journal}
  {Nat. Mater.}\ }\textbf {\bibinfo {volume} {7}},\ \bibinfo {pages} {697}
  (\bibinfo {year} {2008})}\BibitemShut {NoStop}%
\bibitem [{\citenamefont {Wang}\ \emph
  {et~al.}(2012{\natexlab{b}})\citenamefont {Wang}, \citenamefont {Du},
  \citenamefont {Dou},\ and\ \citenamefont {Zhang}}]{TI0}%
  \BibitemOpen
  \bibfield  {author} {\bibinfo {author} {\bibfnamefont {X.}~\bibnamefont
  {Wang}}, \bibinfo {author} {\bibfnamefont {Y.}~\bibnamefont {Du}}, \bibinfo
  {author} {\bibfnamefont {S.}~\bibnamefont {Dou}}, \ and\ \bibinfo {author}
  {\bibfnamefont {C.}~\bibnamefont {Zhang}},\ }\href@noop {} {\bibfield
  {journal} {\bibinfo  {journal} {Phys. Rev. Lett.}\ }\textbf {\bibinfo
  {volume} {108}},\ \bibinfo {pages} {266806} (\bibinfo {year}
  {2012}{\natexlab{b}})}\BibitemShut {NoStop}%
\bibitem [{\citenamefont {Barua}\ \emph {et~al.}(2014)\citenamefont {Barua},
  \citenamefont {Rajeev},\ and\ \citenamefont {Gupta}}]{TI1}%
  \BibitemOpen
  \bibfield  {author} {\bibinfo {author} {\bibfnamefont {S.}~\bibnamefont
  {Barua}}, \bibinfo {author} {\bibfnamefont {K.~P.}\ \bibnamefont {Rajeev}}, \
  and\ \bibinfo {author} {\bibfnamefont {A.~K.}\ \bibnamefont {Gupta}},\
  }\href@noop {} {\bibfield  {journal} {\bibinfo  {journal} {J. Phys. Condens.
  Matter}\ }\textbf {\bibinfo {volume} {27}},\ \bibinfo {pages} {015601}
  (\bibinfo {year} {2014})}\BibitemShut {NoStop}%
\bibitem [{\citenamefont {Hayes}\ \emph {et~al.}(2016)\citenamefont {Hayes},
  \citenamefont {McDonald}, \citenamefont {Breznay}, \citenamefont {Helm},
  \citenamefont {Moll}, \citenamefont {Mark~Wartenbe},\ and\ \citenamefont
  {Analytis}}]{natphy16}%
  \BibitemOpen
  \bibfield  {author} {\bibinfo {author} {\bibfnamefont {I.~M.}\ \bibnamefont
  {Hayes}}, \bibinfo {author} {\bibfnamefont {R.~D.}\ \bibnamefont {McDonald}},
  \bibinfo {author} {\bibfnamefont {N.~P.}\ \bibnamefont {Breznay}}, \bibinfo
  {author} {\bibfnamefont {T.}~\bibnamefont {Helm}}, \bibinfo {author}
  {\bibfnamefont {P.~J.~W.}\ \bibnamefont {Moll}}, \bibinfo {author}
  {\bibfnamefont {A.~S.}\ \bibnamefont {Mark~Wartenbe}}, \ and\ \bibinfo
  {author} {\bibfnamefont {J.~G.}\ \bibnamefont {Analytis}},\ }\href@noop {}
  {\bibfield  {journal} {\bibinfo  {journal} {Nat. Phys.}\ }\textbf {\bibinfo
  {volume} {12}},\ \bibinfo {pages} {916} (\bibinfo {year} {2016})}\BibitemShut
  {NoStop}%
\bibitem [{\citenamefont {Abrikosov}(1999)}]{ab3}%
  \BibitemOpen
  \bibfield  {author} {\bibinfo {author} {\bibfnamefont {A.~A.}\ \bibnamefont
  {Abrikosov}},\ }\href {\doibase 10.1103/PhysRevB.60.4231} {\bibfield
  {journal} {\bibinfo  {journal} {Phys. Rev. B}\ }\textbf {\bibinfo {volume}
  {60}},\ \bibinfo {pages} {4231} (\bibinfo {year} {1999})}\BibitemShut
  {NoStop}%
\bibitem [{\citenamefont {Zhao}\ \emph {et~al.}(2015)\citenamefont {Zhao},
  \citenamefont {Liu}, \citenamefont {Yan}, \citenamefont {An}, \citenamefont
  {Liu}, \citenamefont {Zhang}, \citenamefont {Wang}, \citenamefont {Liu},
  \citenamefont {Jiang}, \citenamefont {Li}, \citenamefont {Wang},
  \citenamefont {Li}, \citenamefont {Mandrus}, \citenamefont {Xie},
  \citenamefont {Pan},\ and\ \citenamefont {Wang}}]{zhaoprb}%
  \BibitemOpen
  \bibfield  {author} {\bibinfo {author} {\bibfnamefont {Y.}~\bibnamefont
  {Zhao}}, \bibinfo {author} {\bibfnamefont {H.}~\bibnamefont {Liu}}, \bibinfo
  {author} {\bibfnamefont {J.}~\bibnamefont {Yan}}, \bibinfo {author}
  {\bibfnamefont {W.}~\bibnamefont {An}}, \bibinfo {author} {\bibfnamefont
  {J.}~\bibnamefont {Liu}}, \bibinfo {author} {\bibfnamefont {X.}~\bibnamefont
  {Zhang}}, \bibinfo {author} {\bibfnamefont {H.}~\bibnamefont {Wang}},
  \bibinfo {author} {\bibfnamefont {Y.}~\bibnamefont {Liu}}, \bibinfo {author}
  {\bibfnamefont {H.}~\bibnamefont {Jiang}}, \bibinfo {author} {\bibfnamefont
  {Q.}~\bibnamefont {Li}}, \bibinfo {author} {\bibfnamefont {Y.}~\bibnamefont
  {Wang}}, \bibinfo {author} {\bibfnamefont {X.-Z.}\ \bibnamefont {Li}},
  \bibinfo {author} {\bibfnamefont {D.}~\bibnamefont {Mandrus}}, \bibinfo
  {author} {\bibfnamefont {X.~C.}\ \bibnamefont {Xie}}, \bibinfo {author}
  {\bibfnamefont {M.}~\bibnamefont {Pan}}, \ and\ \bibinfo {author}
  {\bibfnamefont {J.}~\bibnamefont {Wang}},\ }\href@noop {} {\bibfield
  {journal} {\bibinfo  {journal} {Phys. Rev. B}\ }\textbf {\bibinfo {volume}
  {92}},\ \bibinfo {pages} {041104} (\bibinfo {year} {2015})}\BibitemShut
  {NoStop}%
\bibitem [{\citenamefont {Aamir}\ \emph {et~al.}(2012)\citenamefont {Aamir},
  \citenamefont {Goswami}, \citenamefont {Baenninger}, \citenamefont
  {Tripathi}, \citenamefont {Pepper}, \citenamefont {Farrer}, \citenamefont
  {Ritchie},\ and\ \citenamefont {Ghosh}}]{amirag}%
  \BibitemOpen
  \bibfield  {author} {\bibinfo {author} {\bibfnamefont {M.~A.}\ \bibnamefont
  {Aamir}}, \bibinfo {author} {\bibfnamefont {S.}~\bibnamefont {Goswami}},
  \bibinfo {author} {\bibfnamefont {M.}~\bibnamefont {Baenninger}}, \bibinfo
  {author} {\bibfnamefont {V.}~\bibnamefont {Tripathi}}, \bibinfo {author}
  {\bibfnamefont {M.}~\bibnamefont {Pepper}}, \bibinfo {author} {\bibfnamefont
  {I.}~\bibnamefont {Farrer}}, \bibinfo {author} {\bibfnamefont {D.~A.}\
  \bibnamefont {Ritchie}}, \ and\ \bibinfo {author} {\bibfnamefont
  {A.}~\bibnamefont {Ghosh}},\ }\href@noop {} {\bibfield  {journal} {\bibinfo
  {journal} {Phys. Rev. B}\ }\textbf {\bibinfo {volume} {86}},\ \bibinfo
  {pages} {081203} (\bibinfo {year} {2012})}\BibitemShut {NoStop}%
\bibitem [{\citenamefont {Bud'ko}\ \emph {et~al.}(1998)\citenamefont {Bud'ko},
  \citenamefont {Canfield}, \citenamefont {Mielke},\ and\ \citenamefont
  {Lacerda}}]{redsb}%
  \BibitemOpen
  \bibfield  {author} {\bibinfo {author} {\bibfnamefont {S.~L.}\ \bibnamefont
  {Bud'ko}}, \bibinfo {author} {\bibfnamefont {P.~C.}\ \bibnamefont
  {Canfield}}, \bibinfo {author} {\bibfnamefont {C.~H.}\ \bibnamefont
  {Mielke}}, \ and\ \bibinfo {author} {\bibfnamefont {A.~H.}\ \bibnamefont
  {Lacerda}},\ }\href@noop {} {\bibfield  {journal} {\bibinfo  {journal} {Phys.
  Rev. B}\ }\textbf {\bibinfo {volume} {57}},\ \bibinfo {pages} {13624}
  (\bibinfo {year} {1998})}\BibitemShut {NoStop}%
\bibitem [{\citenamefont {Myers}\ \emph {et~al.}(1999)\citenamefont {Myers},
  \citenamefont {Bud'ko}, \citenamefont {Fisher}, \citenamefont {Islam},
  \citenamefont {Kleinke}, \citenamefont {Lacerda},\ and\ \citenamefont
  {Canfield}}]{reagsb2}%
  \BibitemOpen
  \bibfield  {author} {\bibinfo {author} {\bibfnamefont {K.~D.}\ \bibnamefont
  {Myers}}, \bibinfo {author} {\bibfnamefont {S.~L.}\ \bibnamefont {Bud'ko}},
  \bibinfo {author} {\bibfnamefont {I.~R.}\ \bibnamefont {Fisher}}, \bibinfo
  {author} {\bibfnamefont {Z.}~\bibnamefont {Islam}}, \bibinfo {author}
  {\bibfnamefont {H.}~\bibnamefont {Kleinke}}, \bibinfo {author} {\bibfnamefont
  {A.~H.}\ \bibnamefont {Lacerda}}, \ and\ \bibinfo {author} {\bibfnamefont
  {P.~C.}\ \bibnamefont {Canfield}},\ }\href {\doibase
  https://doi.org/10.1016/S0304-8853(99)00472-2} {\bibfield  {journal}
  {\bibinfo  {journal} {J. Magn. Magn. Mater}\ }\textbf {\bibinfo {volume}
  {205}},\ \bibinfo {pages} {27 } (\bibinfo {year} {1999})}\BibitemShut
  {NoStop}%
\bibitem [{\citenamefont {Shastry}\ and\ \citenamefont
  {Sutherland}(1981)}]{ssl}%
  \BibitemOpen
  \bibfield  {author} {\bibinfo {author} {\bibfnamefont {B.~S.}\ \bibnamefont
  {Shastry}}\ and\ \bibinfo {author} {\bibfnamefont {B.}~\bibnamefont
  {Sutherland}},\ }\href@noop {} {\bibfield  {journal} {\bibinfo  {journal}
  {Physica B+C}\ }\textbf {\bibinfo {volume} {108}},\ \bibinfo {pages} {1069}
  (\bibinfo {year} {1981})}\BibitemShut {NoStop}%
\bibitem [{\citenamefont {Michimura}\ \emph {et~al.}(2009)\citenamefont
  {Michimura}, \citenamefont {Shigekawa}, \citenamefont {Iga}, \citenamefont
  {Takabatake},\ and\ \citenamefont {Ohoyama}}]{michi}%
  \BibitemOpen
  \bibfield  {author} {\bibinfo {author} {\bibfnamefont {S.}~\bibnamefont
  {Michimura}}, \bibinfo {author} {\bibfnamefont {A.}~\bibnamefont
  {Shigekawa}}, \bibinfo {author} {\bibfnamefont {F.}~\bibnamefont {Iga}},
  \bibinfo {author} {\bibfnamefont {T.}~\bibnamefont {Takabatake}}, \ and\
  \bibinfo {author} {\bibfnamefont {K.}~\bibnamefont {Ohoyama}},\ }\href@noop
  {} {\bibfield  {journal} {\bibinfo  {journal} {Jour. of the Phys. Soc. of
  Jpn.}\ }\textbf {\bibinfo {volume} {78}},\ \bibinfo {pages} {024707}
  (\bibinfo {year} {2009})}\BibitemShut {NoStop}%
\bibitem [{\citenamefont {Shin}\ \emph {et~al.}(2017)\citenamefont {Shin},
  \citenamefont {Schlesinger},\ and\ \citenamefont {Shastry}}]{shin}%
  \BibitemOpen
  \bibfield  {author} {\bibinfo {author} {\bibfnamefont {J.}~\bibnamefont
  {Shin}}, \bibinfo {author} {\bibfnamefont {Z.}~\bibnamefont {Schlesinger}}, \
  and\ \bibinfo {author} {\bibfnamefont {B.~S.}\ \bibnamefont {Shastry}},\
  }\href@noop {} {\bibfield  {journal} {\bibinfo  {journal} {Phys. Rev. B}\
  }\textbf {\bibinfo {volume} {95}},\ \bibinfo {pages} {205140} (\bibinfo
  {year} {2017})}\BibitemShut {NoStop}%
\bibitem [{\citenamefont {Momma}\ and\ \citenamefont {Izumi}(2011)}]{vesta}%
  \BibitemOpen
  \bibfield  {author} {\bibinfo {author} {\bibfnamefont {K.}~\bibnamefont
  {Momma}}\ and\ \bibinfo {author} {\bibfnamefont {F.}~\bibnamefont {Izumi}},\
  }\href@noop {} {\bibfield  {journal} {\bibinfo  {journal} {J. Appl. Cryst.}\
  }\textbf {\bibinfo {volume} {44}},\ \bibinfo {pages} {1272} (\bibinfo {year}
  {2011})}\BibitemShut {NoStop}%
\bibitem [{\citenamefont {Siemensmeyer}\ \emph {et~al.}(2008)\citenamefont
  {Siemensmeyer}, \citenamefont {Wulf}, \citenamefont {Mikeska}, \citenamefont
  {Flachbart}, \citenamefont {Gab\'ani}, \citenamefont
  {Mat'a\ifmmode~\check{s}\else \v{s}\fi{}}, \citenamefont {Priputen},
  \citenamefont {Efdokimova},\ and\ \citenamefont {Shitsevalova}}]{t3}%
  \BibitemOpen
  \bibfield  {author} {\bibinfo {author} {\bibfnamefont {K.}~\bibnamefont
  {Siemensmeyer}}, \bibinfo {author} {\bibfnamefont {E.}~\bibnamefont {Wulf}},
  \bibinfo {author} {\bibfnamefont {H.-J.}\ \bibnamefont {Mikeska}}, \bibinfo
  {author} {\bibfnamefont {K.}~\bibnamefont {Flachbart}}, \bibinfo {author}
  {\bibfnamefont {S.}~\bibnamefont {Gab\'ani}}, \bibinfo {author}
  {\bibfnamefont {S.}~\bibnamefont {Mat'a\ifmmode~\check{s}\else \v{s}\fi{}}},
  \bibinfo {author} {\bibfnamefont {P.}~\bibnamefont {Priputen}}, \bibinfo
  {author} {\bibfnamefont {A.}~\bibnamefont {Efdokimova}}, \ and\ \bibinfo
  {author} {\bibfnamefont {N.}~\bibnamefont {Shitsevalova}},\ }\href@noop {}
  {\bibfield  {journal} {\bibinfo  {journal} {Phys. Rev. Lett.}\ }\textbf
  {\bibinfo {volume} {101}},\ \bibinfo {pages} {177201} (\bibinfo {year}
  {2008})}\BibitemShut {NoStop}%
\bibitem [{\citenamefont {{Gab{\'a}ni}}\ \emph {et~al.}(2008)\citenamefont
  {{Gab{\'a}ni}}, \citenamefont {Mat'a\v{s}}, \citenamefont {Priputen},
  \citenamefont {Flachbart}, \citenamefont {Siemensmeyer}, \citenamefont
  {Wulf}, \citenamefont {Evdokimova},\ and\ \citenamefont {Shitsevalova}}]{t2}%
  \BibitemOpen
  \bibfield  {author} {\bibinfo {author} {\bibfnamefont {S.}~\bibnamefont
  {{Gab{\'a}ni}}}, \bibinfo {author} {\bibfnamefont {S.}~\bibnamefont
  {Mat'a\v{s}}}, \bibinfo {author} {\bibfnamefont {P.}~\bibnamefont
  {Priputen}}, \bibinfo {author} {\bibfnamefont {K.}~\bibnamefont {Flachbart}},
  \bibinfo {author} {\bibfnamefont {K.}~\bibnamefont {Siemensmeyer}}, \bibinfo
  {author} {\bibfnamefont {E.}~\bibnamefont {Wulf}}, \bibinfo {author}
  {\bibfnamefont {A.}~\bibnamefont {Evdokimova}}, \ and\ \bibinfo {author}
  {\bibfnamefont {N.}~\bibnamefont {Shitsevalova}},\ }\href@noop {} {\bibfield
  {journal} {\bibinfo  {journal} {Acta. Phys. Pol. A}\ }\textbf {\bibinfo
  {volume} {113}},\ \bibinfo {pages} {227} (\bibinfo {year}
  {2008})}\BibitemShut {NoStop}%
\bibitem [{\citenamefont {Wierschem}\ \emph {et~al.}(2015)\citenamefont
  {Wierschem}, \citenamefont {Sunku}, \citenamefont {Kong}, \citenamefont
  {Ito}, \citenamefont {Canfield}, \citenamefont {Panagopoulos},\ and\
  \citenamefont {Sengupta}}]{t1}%
  \BibitemOpen
  \bibfield  {author} {\bibinfo {author} {\bibfnamefont {K.}~\bibnamefont
  {Wierschem}}, \bibinfo {author} {\bibfnamefont {S.~S.}\ \bibnamefont
  {Sunku}}, \bibinfo {author} {\bibfnamefont {T.}~\bibnamefont {Kong}},
  \bibinfo {author} {\bibfnamefont {T.}~\bibnamefont {Ito}}, \bibinfo {author}
  {\bibfnamefont {P.~C.}\ \bibnamefont {Canfield}}, \bibinfo {author}
  {\bibfnamefont {C.}~\bibnamefont {Panagopoulos}}, \ and\ \bibinfo {author}
  {\bibfnamefont {P.}~\bibnamefont {Sengupta}},\ }\href@noop {} {\bibfield
  {journal} {\bibinfo  {journal} {Phys. Rev. B}\ }\textbf {\bibinfo {volume}
  {92}},\ \bibinfo {pages} {214433} (\bibinfo {year} {2015})}\BibitemShut
  {NoStop}%
\bibitem [{\citenamefont {Sunku}\ \emph {et~al.}(2016)\citenamefont {Sunku},
  \citenamefont {Kong}, \citenamefont {Ito}, \citenamefont {Canfield},
  \citenamefont {Shastry}, \citenamefont {Sengupta},\ and\ \citenamefont
  {Panagopoulos}}]{saiprb}%
  \BibitemOpen
  \bibfield  {author} {\bibinfo {author} {\bibfnamefont {S.~S.}\ \bibnamefont
  {Sunku}}, \bibinfo {author} {\bibfnamefont {T.}~\bibnamefont {Kong}},
  \bibinfo {author} {\bibfnamefont {T.}~\bibnamefont {Ito}}, \bibinfo {author}
  {\bibfnamefont {P.~C.}\ \bibnamefont {Canfield}}, \bibinfo {author}
  {\bibfnamefont {B.~S.}\ \bibnamefont {Shastry}}, \bibinfo {author}
  {\bibfnamefont {P.}~\bibnamefont {Sengupta}}, \ and\ \bibinfo {author}
  {\bibfnamefont {C.}~\bibnamefont {Panagopoulos}},\ }\href@noop {} {\bibfield
  {journal} {\bibinfo  {journal} {Phys. Rev. B}\ }\textbf {\bibinfo {volume}
  {93}},\ \bibinfo {pages} {174408} (\bibinfo {year} {2016})}\BibitemShut
  {NoStop}%
\bibitem [{\citenamefont {Trinh}\ \emph {et~al.}(2018)\citenamefont {Trinh},
  \citenamefont {Mitra}, \citenamefont {Panagopoulos}, \citenamefont {Kong},
  \citenamefont {Canfield},\ and\ \citenamefont {Ramirez}}]{trinprl}%
  \BibitemOpen
  \bibfield  {author} {\bibinfo {author} {\bibfnamefont {J.}~\bibnamefont
  {Trinh}}, \bibinfo {author} {\bibfnamefont {S.}~\bibnamefont {Mitra}},
  \bibinfo {author} {\bibfnamefont {C.}~\bibnamefont {Panagopoulos}}, \bibinfo
  {author} {\bibfnamefont {T.}~\bibnamefont {Kong}}, \bibinfo {author}
  {\bibfnamefont {P.~C.}\ \bibnamefont {Canfield}}, \ and\ \bibinfo {author}
  {\bibfnamefont {A.~P.}\ \bibnamefont {Ramirez}},\ }\href@noop {} {\bibfield
  {journal} {\bibinfo  {journal} {Phys. Rev. Lett.}\ }\textbf {\bibinfo
  {volume} {121}},\ \bibinfo {pages} {167203} (\bibinfo {year}
  {2018})}\BibitemShut {NoStop}%
\bibitem [{\citenamefont {Fisk}\ \emph {et~al.}(1981)\citenamefont {Fisk},
  \citenamefont {Maple}, \citenamefont {Johnston},\ and\ \citenamefont
  {Woolf}}]{t6}%
  \BibitemOpen
  \bibfield  {author} {\bibinfo {author} {\bibfnamefont {Z.}~\bibnamefont
  {Fisk}}, \bibinfo {author} {\bibfnamefont {M.~B.}\ \bibnamefont {Maple}},
  \bibinfo {author} {\bibfnamefont {D.~C.}\ \bibnamefont {Johnston}}, \ and\
  \bibinfo {author} {\bibfnamefont {L.~D.}\ \bibnamefont {Woolf}},\ }\href
  {\doibase 10.1016/0038-1098(81)91111-X} {\bibfield  {journal} {\bibinfo
  {journal} {Solid State Commun.}\ }\textbf {\bibinfo {volume} {39}},\ \bibinfo
  {pages} {1189} (\bibinfo {year} {1981})}\BibitemShut {NoStop}%
\bibitem [{\citenamefont {Yoshii}\ \emph {et~al.}(2006)\citenamefont {Yoshii},
  \citenamefont {Yamamoto}, \citenamefont {Hagiwara}, \citenamefont
  {Shigekawa}, \citenamefont {Michimura}, \citenamefont {Iga}, \citenamefont
  {Takabatake},\ and\ \citenamefont {Kindo}}]{t4}%
  \BibitemOpen
  \bibfield  {author} {\bibinfo {author} {\bibfnamefont {S.}~\bibnamefont
  {Yoshii}}, \bibinfo {author} {\bibfnamefont {T.}~\bibnamefont {Yamamoto}},
  \bibinfo {author} {\bibfnamefont {M.}~\bibnamefont {Hagiwara}}, \bibinfo
  {author} {\bibfnamefont {A.}~\bibnamefont {Shigekawa}}, \bibinfo {author}
  {\bibfnamefont {S.}~\bibnamefont {Michimura}}, \bibinfo {author}
  {\bibfnamefont {F.}~\bibnamefont {Iga}}, \bibinfo {author} {\bibfnamefont
  {T.}~\bibnamefont {Takabatake}}, \ and\ \bibinfo {author} {\bibfnamefont
  {K.}~\bibnamefont {Kindo}},\ }\href@noop {} {\bibfield  {journal} {\bibinfo
  {journal} {J. Phys. Conf. Ser.}\ }\textbf {\bibinfo {volume} {51}},\ \bibinfo
  {pages} {59} (\bibinfo {year} {2006})}\BibitemShut {NoStop}%
\bibitem [{\citenamefont {Suzuki}\ \emph {et~al.}(2010)\citenamefont {Suzuki},
  \citenamefont {Tomita}, \citenamefont {Kawashima},\ and\ \citenamefont
  {Sengupta}}]{prb82}%
  \BibitemOpen
  \bibfield  {author} {\bibinfo {author} {\bibfnamefont {T.}~\bibnamefont
  {Suzuki}}, \bibinfo {author} {\bibfnamefont {Y.}~\bibnamefont {Tomita}},
  \bibinfo {author} {\bibfnamefont {N.}~\bibnamefont {Kawashima}}, \ and\
  \bibinfo {author} {\bibfnamefont {P.}~\bibnamefont {Sengupta}},\ }\href
  {\doibase 10.1103/PhysRevB.82.214404} {\bibfield  {journal} {\bibinfo
  {journal} {Phys. Rev. B}\ }\textbf {\bibinfo {volume} {82}},\ \bibinfo
  {pages} {214404} (\bibinfo {year} {2010})}\BibitemShut {NoStop}%
\bibitem [{\citenamefont {Dublenych}(2012)}]{t5}%
  \BibitemOpen
  \bibfield  {author} {\bibinfo {author} {\bibfnamefont {Y.~I.}\ \bibnamefont
  {Dublenych}},\ }\href@noop {} {\bibfield  {journal} {\bibinfo  {journal}
  {Phys. Rev. Lett.}\ }\textbf {\bibinfo {volume} {109}},\ \bibinfo {pages}
  {167202} (\bibinfo {year} {2012})}\BibitemShut {NoStop}%
\bibitem [{\citenamefont {Ye}\ \emph {et~al.}(2017)\citenamefont {Ye},
  \citenamefont {Suzuki},\ and\ \citenamefont {Checkelsky}}]{chkprb}%
  \BibitemOpen
  \bibfield  {author} {\bibinfo {author} {\bibfnamefont {L.}~\bibnamefont
  {Ye}}, \bibinfo {author} {\bibfnamefont {T.}~\bibnamefont {Suzuki}}, \ and\
  \bibinfo {author} {\bibfnamefont {J.~G.}\ \bibnamefont {Checkelsky}},\
  }\href@noop {} {\bibfield  {journal} {\bibinfo  {journal} {Phys. Rev. B}\
  }\textbf {\bibinfo {volume} {95}},\ \bibinfo {pages} {174405} (\bibinfo
  {year} {2017})}\BibitemShut {NoStop}%
\bibitem [{\citenamefont {Shekhar}\ \emph {et~al.}(2015)\citenamefont
  {Shekhar}, \citenamefont {Nayak}, \citenamefont {Sun}, \citenamefont
  {Schmidt}, \citenamefont {Nicklas}, \citenamefont {Leermakers}, \citenamefont
  {Zeitler}, \citenamefont {Skourski}, \citenamefont {Wosnitza}, \citenamefont
  {Liu}, \citenamefont {Chen}, \citenamefont {Schnelle}, \citenamefont
  {Borrmann}, \citenamefont {Grin}, \citenamefont {Felser},\ and\ \citenamefont
  {Yan}}]{NbP}%
  \BibitemOpen
  \bibfield  {author} {\bibinfo {author} {\bibfnamefont {C.}~\bibnamefont
  {Shekhar}}, \bibinfo {author} {\bibfnamefont {A.~K.}\ \bibnamefont {Nayak}},
  \bibinfo {author} {\bibfnamefont {Y.}~\bibnamefont {Sun}}, \bibinfo {author}
  {\bibfnamefont {M.}~\bibnamefont {Schmidt}}, \bibinfo {author} {\bibfnamefont
  {M.}~\bibnamefont {Nicklas}}, \bibinfo {author} {\bibfnamefont
  {I.}~\bibnamefont {Leermakers}}, \bibinfo {author} {\bibfnamefont
  {U.}~\bibnamefont {Zeitler}}, \bibinfo {author} {\bibfnamefont
  {Y.}~\bibnamefont {Skourski}}, \bibinfo {author} {\bibfnamefont
  {J.}~\bibnamefont {Wosnitza}}, \bibinfo {author} {\bibfnamefont
  {Z.}~\bibnamefont {Liu}}, \bibinfo {author} {\bibfnamefont {Y.}~\bibnamefont
  {Chen}}, \bibinfo {author} {\bibfnamefont {W.}~\bibnamefont {Schnelle}},
  \bibinfo {author} {\bibfnamefont {H.}~\bibnamefont {Borrmann}}, \bibinfo
  {author} {\bibfnamefont {Y.}~\bibnamefont {Grin}}, \bibinfo {author}
  {\bibfnamefont {C.}~\bibnamefont {Felser}}, \ and\ \bibinfo {author}
  {\bibfnamefont {B.}~\bibnamefont {Yan}},\ }\href@noop {} {\bibfield
  {journal} {\bibinfo  {journal} {Nat. Phys.}\ }\textbf {\bibinfo {volume}
  {11}},\ \bibinfo {pages} {645} (\bibinfo {year} {2015})}\BibitemShut
  {NoStop}%
\bibitem [{\citenamefont {Goodings}(1963)}]{goodings}%
  \BibitemOpen
  \bibfield  {author} {\bibinfo {author} {\bibfnamefont {D.~A.}\ \bibnamefont
  {Goodings}},\ }\href {\doibase 10.1103/PhysRev.132.542} {\bibfield  {journal}
  {\bibinfo  {journal} {Phys. Rev.}\ }\textbf {\bibinfo {volume} {132}},\
  \bibinfo {pages} {542} (\bibinfo {year} {1963})}\BibitemShut {NoStop}%
\bibitem [{\citenamefont {Madduri}\ and\ \citenamefont {Kaul}(2017)}]{kaul}%
  \BibitemOpen
  \bibfield  {author} {\bibinfo {author} {\bibfnamefont {P.~V.~P.}\
  \bibnamefont {Madduri}}\ and\ \bibinfo {author} {\bibfnamefont {S.~N.}\
  \bibnamefont {Kaul}},\ }\href@noop {} {\bibfield  {journal} {\bibinfo
  {journal} {Phys. Rev. B}\ }\textbf {\bibinfo {volume} {95}},\ \bibinfo
  {pages} {184402} (\bibinfo {year} {2017})}\BibitemShut {NoStop}%
\bibitem [{\citenamefont {Lin}\ \emph {et~al.}(2015)\citenamefont {Lin},
  \citenamefont {Fauqu{\'e}},\ and\ \citenamefont {Behnia}}]{science}%
  \BibitemOpen
  \bibfield  {author} {\bibinfo {author} {\bibfnamefont {X.}~\bibnamefont
  {Lin}}, \bibinfo {author} {\bibfnamefont {B.}~\bibnamefont {Fauqu{\'e}}}, \
  and\ \bibinfo {author} {\bibfnamefont {K.}~\bibnamefont {Behnia}},\
  }\href@noop {} {\bibfield  {journal} {\bibinfo  {journal} {Science}\ }\textbf
  {\bibinfo {volume} {349}},\ \bibinfo {pages} {945} (\bibinfo {year}
  {2015})}\BibitemShut {NoStop}%
\bibitem [{\citenamefont {Abrikosov}(2000)}]{ab4}%
  \BibitemOpen
  \bibfield  {author} {\bibinfo {author} {\bibfnamefont {A.~A.}\ \bibnamefont
  {Abrikosov}},\ }\href@noop {} {\bibfield  {journal} {\bibinfo  {journal}
  {Europhys. Lett.}\ }\textbf {\bibinfo {volume} {49}},\ \bibinfo {pages} {789}
  (\bibinfo {year} {2000})}\BibitemShut {NoStop}%
\bibitem [{Note1()}]{Note1}%
  \BibitemOpen
  \bibinfo {note} {For calculation details please see Ref.~\protect \citenum
  {shin}}\BibitemShut {NoStop}%
\bibitem [{Note2()}]{Note2}%
  \BibitemOpen
  \bibinfo {note} {Above this \protect \textit {T}-value, we previously
  observed a quadratic \protect \textit {MR} for $B\parallel c$-axis. Please
  see ~Ref.\protect \citenum {saiprb} for details}\BibitemShut {NoStop}%
\bibitem [{Note3()}]{Note3}%
  \BibitemOpen
  \bibinfo {note} {Estimating $n_e^{\protect \text {pocket}}$ is tricky, since
  the pocket is reduced to a point in the zero-field first principle
  calculations. However, as a rough estimate, we can use the calculated value
  for the other smallest Fermi pockets\cite {shin}, viz., $n_e ~(\sim \SI
  {e24}{\meter ^{-3}}$.)}\BibitemShut {NoStop}%
\bibitem [{Note4()}]{Note4}%
  \BibitemOpen
  \bibinfo {note} {A comparison between the residual resistivity (\protect
  \textit {RR}) of our sample and similarly grown \ce {TmB4} crystals with
  known impurity density\cite {jjap} allows us to estimate $N_i$ (assuming the
  \protect \textit {RR} arises solely from scattering of electrons off
  impurity), as $\SI {e21}{\meter ^{-3}}$. For details, see Supplementary
  information}\BibitemShut {NoStop}%
\bibitem [{\citenamefont {Okada}\ \emph {et~al.}(1994)\citenamefont {Okada},
  \citenamefont {Kudou}, \citenamefont {Yu},\ and\ \citenamefont
  {Lundström}}]{jjap}%
  \BibitemOpen
  \bibfield  {author} {\bibinfo {author} {\bibfnamefont {S.}~\bibnamefont
  {Okada}}, \bibinfo {author} {\bibfnamefont {K.}~\bibnamefont {Kudou}},
  \bibinfo {author} {\bibfnamefont {Y.}~\bibnamefont {Yu}}, \ and\ \bibinfo
  {author} {\bibfnamefont {T.}~\bibnamefont {Lundström}},\ }\href@noop {}
  {\bibfield  {journal} {\bibinfo  {journal} {Jpn. J. Appl. Phys}\ }\textbf
  {\bibinfo {volume} {33}},\ \bibinfo {pages} {2663} (\bibinfo {year}
  {1994})}\BibitemShut {NoStop}%
\end{thebibliography}%
%


\end{document}